\documentclass{article}

\usepackage{arxiv}

\usepackage[utf8]{inputenc} 
\usepackage[T1]{fontenc}    
\usepackage{hyperref}       
\usepackage{url}            
\usepackage{booktabs}       
\usepackage{amsfonts}       
\usepackage{nicefrac}       
\usepackage{microtype}      
\usepackage{graphicx}
\usepackage{doi}
\usepackage{amsmath,amssymb,amsfonts}
\usepackage{algorithmic}
\usepackage{textcomp}
\usepackage{adjustbox}
\usepackage{threeparttable}
\usepackage{booktabs}
\usepackage{multirow}
\usepackage{authblk}
\usepackage{tabularx}
\usepackage{subfigure}

\begin{document}

\title{Identifying Autism Spectrum Disorder Based on Individual-Aware Down-Sampling and Multi-Modal Learning}

\date{} 					

\author[1]{\small Li Pan}
\author[2]{\small Jundong Liu}
\author[3]{\small Mingqin Shi}
\author[4]{\small Chi Wah Wong}
\author[1,2,5]{\small  Kei Hang Katie Chan}

\affil[1]{\footnotesize Department of Electrical Engineering, City University of Hong Kong, Hong Kong SAR}
\affil[2]{\footnotesize Department of Biomedical Sciences, City University of Hong Kong, Hong Kong SAR}
\affil[3]{\footnotesize School of Basic Medical Sciences, Yunnan University of Chinese Medicine, Kunming, China}
\affil[4]{\footnotesize Department of Applied AI and Data Science, City of Hope, Duarte CA 91010, USA}
\affil[5]{\footnotesize Department of Epidemiology, Brown University, Providence RI 02912, USA}


\renewcommand{\shorttitle}{Identifying ASD Based on Individual-Aware Down-Sampling and Multi-Modal Learning}


\maketitle

\begin{abstract}
Autism Spectrum Disorder(ASD) is a set of neurodevelopmental conditions that affect patients' social abilities. In recent years, many studies have employed deep learning to diagnose this brain dysfunction through functional MRI (fMRI). However, existing approaches solely focused on the abnormal brain functional connections but ignored the impact of regional activities. Due to this biased prior knowledge, previous diagnosis models suffered from inter-site measurement heterogeneity and inter-individual phenotypic differences. To address this issue, we propose a novel feature extraction method for fMRI that can learn a personalized lower-resolution representation of the entire brain networking regarding both the functional connections and regional activities. Specifically, we abstract the brain imaging as a graph structure and straightforwardly downsample it to substructures by hierarchical graph pooling. To further recalibrate the distribution of the extracted features under phenotypic information, we subsequently embed the sparse feature vectors into a population graph, where the hidden inter-subject heterogeneity and homogeneity are explicitly expressed as inter- and intra-community connectivity differences, and utilize Graph Convolutional Networks to learn the node embeddings. By these means, our framework can extract features directly and efficiently from the entire fMRI and be aware of implicit inter-individual variance. We have evaluated our framework on the ABIDE-I dataset with 10-fold cross-validation. The present model has achieved a mean classification accuracy of 87.62\% and a mean AUC of 0.92, better than the state-of-the-art methods. Our code and support documents are available at \underline{github.com/jhonP-Li/ASD\_GP\_GCN}.
\end{abstract}


\section{Introduction}
\label{sec:introduction}
Autism spectrum disorder (ASD), a range of brain developmental disorders, has commonly been studied worldwide. In 2020, a survey reported that approximately 1 in 45 children in the U.S. was diagnosed with this disease caused by both genetic and environmental factors\cite{maenner2020prevalence}. This mental disorganization, which will result in difficulties with social interaction and communication, can be noticed at the early age of a child. However, another study, conducted in the U.K., showed that the current time-consuming diagnosis process could lead to a delay of around 3.5 years from the point at which parents first consult a doctor to the confirmation of an ASD diagnosis\cite{crane2016experiences}, which results in unnecessary panic and late intervention. 

Similar to physical disease diagnosis, this brain dysfunction can be detected with pathological manifestations. In \cite{carmody2010regional}, the authors discovered that there existed structural differences in certain areas between patients suffering from autism and control subjects. \cite{villalobos2005reduced} reported that abnormal brain function connections were found in ASD subjects. However, evaluated only on a few samples, these diagnosis methods cannot be generalized. In recent years, brain imaging analysis based on deep learning and machine learning, tested on large datasets, has been widely studied. \cite{dvornek2017identifying} employed Long Short-Term Memory (LSTM) to analyze the time-series data of fMRI automatically. \cite{heinsfeld2018identification} applied Deep Neural Networks (DNN) directly on the fMRI and reported the performance improvement compared to Support Vector Machine (SVM) and Random Forest (RF). In \cite{sherkatghanad2020automated}, the authors designed a Convolutional Neural Networks (CNN) architecture with fMRI as input and achieved slightly better performance than the DNN did. \cite{khosla20183d} reached the best performance of end-to-end CNN models by 3D-CNN and the ensemble brain atlas. 

Limited by the ability to process structured brain networking, the naive end-to-end implementations of CNN models have reached the bottleneck. On the one side, the functional connections of brain regions do not follow the spatial distributions of the areas, e.g., a node may interact with another far away from it \cite{parisot2018disease}. However, the convolution kernels can only extract features from spatial neighborhoods for each pixel. To be aware of those cross-space connections, CNN needs more convolution layers to form a wider receptive field, which backfires to overfitting due to the lack of samples. On the other side, to downsample the raw inputs, many methods have focused on selecting a certain number of functional connections. These methods represent brain imaging as a correlation coefficient matrix of which the elements denote the covariances between every two regions based on their time-series signals \cite{nielsen2013multisite, abraham2017deriving, kazeminejad2020importance,mostafa2019diagnosis, parisot2018disease, wang2020aimafe, liu2020improved}. Specifically, in \cite{mostafa2019diagnosis}, the authors elaborately constructed a workflow to extract features from functional connections and achieved state-of-the-art performance on the ABIDE I dataset using a linear classifier. However, all the existing feature selection methods solely extract features from the pairwise regional correlation matrix. This inflexible Euclidean representation of brain imaging did not only ignore the details of regional signals, which are believed to relate to ASD \cite{saenz2020adhd,harris2006brain, hadjikhani2007abnormal, kim2015abnormal}, but also omit the latent interaction among the connections.

To leverage the functional connections and regional activities, the brain imaging needs to be described as a graph structure, which perspicuously expresses the functional interactivity among regions as edges among nodes. Some studies have employed this non-Euclidean form to simulate brain networking and discover group-level brain biomarkers of ASD \cite{ktena2018metric, li2019graph, yao2019triplet, rakhimberdina2019linear,li2020pooling}, e.g., \cite{ktena2018metric} constructed a personalized brain connectivity graph for each individual and measured the inter-individual graph structural difference using graph convolutional networks. \cite{li2019graph} utilized graph convolutional networks on the geometric representation of the brain and tried to interpret the learned connections as ASD biomarkers. To find the abnormal brain networks that may interact with ASD, those graph convolutional methods have barely downscaled the input brain imaging. In other words, they focused on directly analyzing the graph-level information but did not extract higher-order features from it like the above selection methods, which makes the methods susceptible to inter-individual differences and even temporary brain activities \cite{baliki2006chronic, tavor2016task}. Hence, limited by the surfeit model complexity and individual brain differences, the ASD identification accuracy of those models is not promising, making the inference of biomarkers unconvincing.

In the context of this ASD disease prediction problem, another challenge is the non-imaging difference between individuals, i.e., gender, handedness, IQ, etc. Though this information is not present in the fMRI, it does affect the probability of suffering from ASD. For example, \cite{loomes2017male} indicated that One in every 42 males and one in 189 females in the United States is diagnosed with an autism spectrum disorder. \cite{rysstad2018there} reported the correlation between the handedness and ASD. Besides, the fMRI scanning devices and measurement parameters from different data collection sites are not strictly the same \cite{mostafa2019diagnosis}. Those hidden factors have caused the non-identity distribution of features and lowered the generalization ability of models. To address this, some authors manually forced those settings to be the same by hard clustering strategy on samples. For example, in \cite{guo2017diagnosing, kong2019classification, li2018novel, bi2018diagnosis}, the authors elaborately selected training and testing samples from a certain data collection site. Hence, the implicit differences among samples were further narrowed, and these methods achieved much better classification performance than the models evaluated on the entire dataset. Although this hard clustering strategy did prove the feasibility of ASD diagnosis based on deep learning, the generalization ability of the models cannot be guaranteed as the number of samples has been dramatically reduced in that way.

To address the above issue, Graph Convolutional Networks (GCN) can be adopted to recalibrate the features extracted from brain imaging, according to the non-imaging data. Unlike assigning each subject into a cluster, we embed each subject into a population graph, where nodes denote individuals and edges represent the phenotypic similarity between every two nodes. Thus, the hidden inter-subject heterogeneity and homogeneity are explicitly expressed as inter- and intra-community connectivity differences. Some methods succeeded in fusing imaging and text information in this way but achieved lower classification accuracy than unimodal methods owing to inefficient brain imaging feature extraction \cite{parisot2017spectral, arya2020fusing, rakhimberdina2019linear}. Namely, in \cite{arya2020fusing}, the authors employed 3D CNN to extract features from fMRI and Variational Autoencoder to extract features from MRI, which are not efficient as previously discussed. Although this framework has considered functional, structural, and phenotypic information, it achieved lower performance than the method that directly utilized 3D CNN on fMRI \cite{khosla20183d}. 

\begin{table*}[t!]
\caption{Overview of the ABIDE I dataset preprocessed by CPAC}
\centering
\label{table:data}
\begin{adjustbox}{center}
\begin{threeparttable}

\begin{tabular*}{\linewidth}{@{\extracolsep{\fill}}lccc cc c c cc c}
\toprule
\multirow{2}{*}{Sites} & \multicolumn{2}{c}{Age(year)} & \multicolumn{2}{c}{Gender} & \multicolumn{4}{c}{handedness} & \multicolumn{2}{c}{Diagnostic group}\\
\cmidrule{2-3} \cmidrule{4-5} \cmidrule{6-9} \cmidrule{10-11}
  & Min & Max & Male & Female & Left & Right & Ambi$^*$ & Mixed & ASD & Control  \\
\midrule
CALTECH	 & 17.0 & 56.2 & 10 & 5 & 1& 13& 1& 0 & 5 & 10\\
CMU	 & 19.0 & 33.0 & 7 & 4 & 1& 10 & 0 & 0 & 6 & 5\\
KKI	 & 8.2 & 12.8 & 24 & 9 & 1& 27 & 0 & 5 & 12 & 21\\
LEUVEN	 & 12.1 & 32.0 & 49 & 7 & 7 & 49 & 0 & 0 & 26 & 30\\
MAX\_MUN	 & 7.0 & 58.0 & 42 & 4 & 2 & 44 & 0 & 0 & 19 & 27\\
NYU	 & 6.5 & 39.1 & 136 & 36 & \multicolumn{4}{c}{N/A} & 74 & 98\\
OHSU	 & 8.0 & 15.2 & 25 & 0 & 1 & 24 & 0 & 0 & 12 & 13\\
OLIN	 & 10.0 & 24.0 & 23 & 5 & 5 & 23 & 0 & 0 & 14 & 14\\
PITT$^\dagger$	 & 9.3 & 35.2 & 43 & 7 & 4 & 45 & 0 & 0 & 24 & 26\\
SBL$^\dagger$	 & 20.0 & 49.0 & 26 & 0 & 1 & 0 & 0 & 0 & 12 & 14\\
SDSU	 & 8.7 & 17.2 & 21 & 6 & 2 & 25 & 0 & 0 & 8 & 19\\
STANFORD	 & 7.5 & 12.9 & 18 & 7 & 3 & 20 & 2 & 0 & 12 & 13\\
TRINITY	 & 12.0 & 25.7 & 44 & 0 & 0 & 44 & 0 & 0 & 19 & 25\\
UCLA	 & 8.4 & 17.9 & 74 & 11 & 9 & 76 & 0 & 0 & 48 & 37\\
UM$^\dagger$ & 8.2 & 28.8 & 93 & 27 & 15 & 97 & 0 & 0 & 47 & 73\\
USM & 8.8 & 50.2 & 67 & 0 & \multicolumn{4}{c}{N/A} & 43 & 24\\
YALE & 7.0 & 17.8 & 25 & 16 & 7 & 34 & 0 & 0 & 22 & 19\\
\cmidrule{1-11}
Total & 6.5 & 58.0 & 727 & 144 & 59 & 531 & 3 & 5 & 403 & 468\\
\bottomrule
\end{tabular*}

\begin{tablenotes}
\footnotesize
\item[*] Ambi: Ambidextrous
\item[$^\dagger$] The handedness information of some subjects is unavailable.
\end{tablenotes}

\end{threeparttable}
\end{adjustbox}
\end{table*} 

In this study, we propose a novel framework that incorporates self-attention graph pooling and graph convolutional networks. We explicit graph pooling to downsample the structured form of brain imaging, whereas previous brain modeling methods can solely extract features from functional connections. Unlike existing graph-level analysis, we individually downsample and flatten the brain imaging to sparse vectors, and implement Multilayer Perceptron to extract higher-level information from the selected subgraphs. To fuse the imaging and non-imaging information, we then initialize a population graph, where nodes represent individuals and edges denote phenotypic similarities. Assigning each subject with the extracted brain imaging features, we employ Graph Convolutional Networks to learn the node embeddings on the population graph, which succeed in regularizing the individual features according to phenotypic property. Having merged functional connections, regional activities, and non-imaging data, our framework presents its superior in ASD diagnosis, reaching an accuracy of 87.62\% on ABIDE I dataset. The main contributions of our work are four-fold:

\begin{itemize}
\item[1)] We have developed a self-attention feature extraction method on fMRI feature extraction tasks. The unsupervised graph pooling efficiently extracts features from the non-Euclidean brain networking. By weighing both the functional connections and regional signals simultaneously, this downsampling method retains the crucial information for diagnostic classification.
\item[2)] We have stabilized the model performance by fusing the multi-modal information through graph convolutional networks and intuitively illustrated its efficiency.
\item[3)] Different from hard selecting universal biomarkers of ASD, our framework tends to select a personalized substructure of brain networking for each individual. This novel strategy has detected inter-group heterogeneity and intra-group homogeneity regarding brain activities.
\item[4)] We have constructed an ASD diagnosing framework, which outperforms state-of-the-art methods on the ABIDE I dataset, reaching a classification accuracy of 87.62\%. This clinically meaningful method could contribute to early detection and intervention for ASD.

\end{itemize} 

The rest of the paper is organized as follows: Section II introduces the datasets and illustrates the details of the proposed model. In Section III, we present the experimental setup, evaluation metrics, experimental results and comparison with other methods, ablation study, and intuitive exhibition of model mechanisms. Finally, we draw the conclusion in Section IV.

\section{Materials and Methods}
\subsection{ABIDE Dataset} \label{subsection: abide}
Constructed by \cite{di2014autism}, the ABIDE I dataset contains a variety of information of 1,112 subjects, i.e., MRI. fMRI, and phenotype data, collected from 17 international sites. To reduce the fMRI measurement error, current studies are focusing on the preprocessed data. \cite{craddock2013neuro} performed four different preprocessing pipelines on the original material. To compare with other methods \cite{parisot2018disease, mostafa2019diagnosis, heinsfeld2018identification, khosla20183d, nielsen2013multisite}, in the current paper, we have used the data preprocessed by the Configurable Pipeline for the Analysis of Connectomes (CPAC). Built by \cite{craddock2013neuro}, the chosen functional preprocessing pipeline includes time slicing, motion correction, skull-stripping, global mean intensity normalization, and nuisance signal regression. Thus, the noise caused by unrelated motions, like the heart beating, is reduced. To further regularize the input sample features, we have also employed band-pass filtering and global signal regression.

As shown in the table \ref{table:data}, the processed data contains 403 ASD subjects and 468 typical control subjects. Caused by the measurement difference among different sites, some categories of the phenotype data are not or partially collected, like handedness information. Moreover, the distribution of the dataset is unbalanced on some features. For example, the 17 sites only have collected 144 female samples but collected 727 male samples. According to the view of \cite{loomes2017male}, gender is a rather important factor affecting the probability of ASD. This unbalanced feature distribution, which is not present in the MRI or fMRI data, has caused the non-identical distribution of features and thus affected the performance of unimodal learning models.

\begin{figure*}[h!]
\centering
\includegraphics[width=\textwidth]{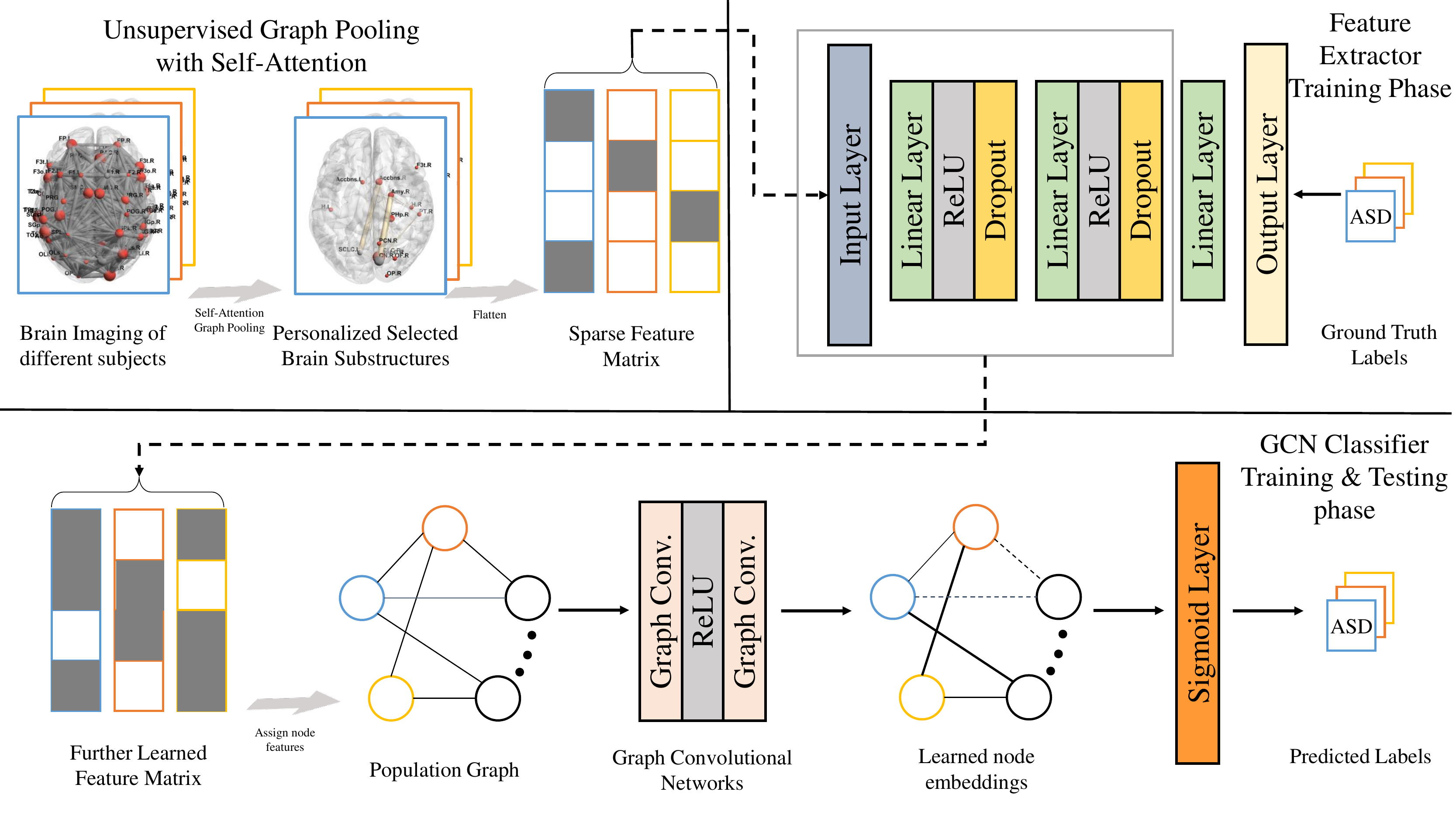}
\caption{Overview of the proposed framework; The top-left part illustrates self-attention graph pooling in section \ref{sec:gp}. The top-right part indicates the training phase of feature extractor, where we train the MLP using the pooling output. In the bottom part, we construct a population graph, where nodes denote subjects and edges represent interindividual phenotypic similarity. We then train a GCN model by assigning each node with dense features extracted by MLP.}
\label{fig:model}
\end{figure*}  

To further reduce the dimensionality of input data, fMRI is separated logically as signals of regions of interest (ROIs). The voxel-wise time series is thus paraphrased as the time series of regional signals. Proposed by \cite{desikan2006automated, goldstein2007hypothalamic, makris2006decreased, frazier2005structural}, the Harvard-Oxford atlas is split into cortical and subcortical structural probabilistic atlases. The HO atlas, which has also been selected by other works\cite{khosla20183d, parisot2018disease, heinsfeld2018identification}, is filtered with a 25$\%$ threshold and subsequently divided into left and right hemispheres at the midline. The ROIs represent 110 functional brain regions, i.e., left and right Hippocampus, left and right Cuneal Cortex, left and right Planum Temporale, left and right Occipital Pole, etc. Thus, the original 4D brain imaging is further downsampled to a 2D data structure, containing 110 regions and the corresponding time series for each area.

\subsection{Model Overview}
As shown in the figure \ref{fig:model}, the whole pipeline consists of three main parts. First, the unsupervised graph pooling directly downsamples the graph representation of the brain to a sparse brain networking. We then train a multilayer perception using the flattened features and ground truth labels to extract higher-order features from the pooling results. Finally, we employ a two-layer graph convolutional network to learn the node embeddings by embedding every individual into the population graph and building edges according to the phenotypic information.

\subsection{Graph Pooling}\label{sec:gp}
In this section, we develop a self-attention unsupervised graph pooling strategy inspired by \cite{zhang2019hierarchical} to select the crucial subgraphs of brain networking for each subject individually. It improves the existing brain imaging feature extractors in two main aspects. First, it can extract features directly from the entire graph structure, while other methods only extract features from functional connections. Second, this graph pooling operation downsamples graph without supervision. In other words, this step can be directly added to other related frameworks without any additional training cost. Graph pooling, as a downsampling method for graph structure, is a central component of graph convolutional networks in graph classification tasks. For example, the intuitive idea is to average all node embeddings to represent the entire graph \cite{duvenaud2015convolutional}. Compared to other graph pooling methods\cite{ying2018hierarchical, gao2019graph, lee2019self}, the proposed graph pooling procedure is designed to preserve the information and connectivity of the graph and reduce the information redundancy. As shown in figure \ref{fig:ratio}, this strategy can automatically select key brain subgraphs, which are sufficient for high-accuracy ASD/TC classification.

Before this implementation of graph pooling, other models trained classifiers, like MLP, or constructed a specific feature extracting framework to extract features from the functional connections\cite{parisot2018disease, wang2020aimafe, liu2020improved, mostafa2019diagnosis} which are represented as the correlation coefficients of every two regions. At this phase, those methods have directly ignored the details of regional brain activities. This intuitive method, mapping the two vectors to a float ranging from -1 to 1, has reduced the data complexity. However,  regarding the time series of brain regions as node features and functional connections as edges, this edge-only feature extraction strategy has caused much more information loss to the entire graph structure. Moreover, studies have proved the importance of the node features in the diagnosis of ASD. In \cite{saenz2020adhd,harris2006brain, hadjikhani2007abnormal, kim2015abnormal}, the authors reported the abnormal regional activities among ASD subjects. Thus, leveraging the information of both functional connections and regional activities, graph pooling has proved its superior in this brain disorder diagnosis as shown in section \ref{sec:ipr}.

\subsubsection{Graph Representation of Brain Imaging}
After being labeled as 110 regions according to the HO atlases, fMRI can be abstracted as a graph structure, where nodes denote brain regions and edges indicate functional connections. Initially, every node is assigned with a feature vector that represents the time series of regional activity. In \cite{parisot2018disease}, the authors defined 6105 brain functional connections by connecting right-side regions to the left side and left-side regions to the right. Inspired by it, we construct a graph representation of brain imaging, where regions are connected according to the same strategy and all regions are connected with the global mean time series to reduce measurement error further. In short, the input brain graph structure contains 111 nodes and 6215 edges.

\subsubsection{Node Selection}
\label{sec:hgp}
The self-attention graph pooling consists of two parts: First, it selects the nodes based on the criterion of minimizing graph information loss. Subsequently, to connect the probably isolated subgraph caused by the node selection and recorrect the initial brain regional connections to some degree, an unsupervised edge prediction method is employed between the two-hop neighbors of each node and itself. At the first component, a node information score is defined as the $\mathcal{L}_1$ norm of the Manhattan distance between the node features itself and the one constructed from its neighbors \cite{zhang2019hierarchical}:
\begin{equation}
S \ = \ \gamma(g) \ = \ \left\| (I - (D^{(l)})^{-1}A^{(l)})H^{(l)} \right\|_1 
\end{equation}
where  $A^{(l)}$ and $H^{(l)}$ are the adjacency and node features matrices of the $l$-th layer. The information of edges is present implicitly as the connections among nodes. $I$ represents the identity matrix and $D^{(l)}$ denotes the $l$-th layer diagonal degree matrix of $A^{(l)}$. $\left\|\cdot\right\|_1 $ performs the $\mathcal{L}_1$ norm row-wisely. The vector $S$ contains the information score of each node, which indicates its importance at this selection stage. The nodes are then selected by ranking and selecting the top-K ones regarding the information score:

\begin{equation}
\begin{gathered}
idx  \ = \ top( S, \lceil r * n^{(l)}\rceil )\\
H^{(l+1)} \ = \ H^{(l)}(idx,:) \\
A^{(l+1)} \ = \ A^{(l)}(idx, idx) \\
\end{gathered}
\end{equation}
where $r$ is the pooling ratio which is set manually and will be discussed in the section \ref{sec:ipr}. The function $top(\cdot)$ returns the indices of top $n^{(l+1)} = \lceil r * n^{(l)}\rceil$ values of the information scores $S$. $H^{(l)}(idx,:)$ and $A^{(l)}(idx, idx)$ performs the element selection according to the indices of top information scores. Thus, in the $l$-th layer, $n^{(l+1)}$ nodes are remained and others are removed.

Intuitively, the information score of a node is the feature difference between the average value of its neighbors and itself. The greater the difference, the higher the information score, and the less likely to remove the node. For example, if the feature of a node is equal to the average feature of its neighbors, it may be safe to drop this node without further information loss to the entire graph. On the other side, this selection method simulates a probable universal strategy for removing the information redundancy of fMRI: If the blood oxygen level activity is close to its neighbors, the area may be regarded as coactivated with its neighbors. After Removing all those nodes, the remaining ones may be activated in the first order, which may act like a trigger that has launched the sequence of regional brain activities.

\subsubsection{Edge Prediction}
At the same time, the node selection method may isolate some subgraphs and be susceptible to the initialization of brain graph structure, as it can not learn new connections beyond the given ones. To preserve the completeness of the subgraph, \cite{zhang2019hierarchical} developed a differentiable edge detection method based on the node features, which involved superfluous training overhead. Instead, we have designed a self-attention approach to predict underlying links among selected nodes:

\begin{equation}
E^{(l)}(p, q) \ = \ \frac{H^{(l)}(p,:) \cdot H^{(l)}(q,:)}{\left\|H^{(l)}(p,:)\right\| \left\|H^{(l)}(q,:)\right\|} + A^{(l)}(p,q) \\
\end{equation}
where $E^{(l)}(p, q)$ represents the similarity score between the two nodes, $ H^{(l)}$ denotes the feature matrix at the $l$-th layer. The corresponding element of adjacency matrix $A^{(l)}(p,q)$ is added to the cosine similarity of the two feature vectors to assign a more significant similarity score to the directly connected nodes. The similarity score is then normalized by the sparse attention mechanism proposed by \cite{ martins2016softmax}:
\begin{equation}
\begin{gathered}
Sim^{(l)}(p, :) \ := \ \mathop{argmin}_{\bold{n} \in \Delta^{K-1}} \left\| \bold{n} - E^{(l)}(p,:) \right\|^2\\
where \ \ \Delta^{K-1} \ = \ \{\bold{n} \in \mathbb{R}^K \ | \  1^T\bold{n}=1, \bold{n}\ge0\}\\
\end{gathered}
\end{equation}
where $\Delta^{K-1}$ is a (K-1)-dimensional probability simplex and K denotes the number of nodes in the brain graph. For arbitrary node $p$, the softmax function normalizes the similarity scores between it and other nodes to probability distributions in $\Delta^{K-1}$. However, this probabilistic approach retains small non-zero values of normalized similarities, which increases the complexity of downsampled subgraph.  \cite{ martins2016softmax} projects the target vector onto simplex $\Delta^{K-1}$ and achieves sparsity when hitting the boundary. Specifically, the adjacency matrix is updated as the optimum of this quadratically constrained optimization problem. The closed-form solution is as follows:
\begin{equation}
A^{(l+1)}(p,q) \ = \ \left[E^{(l)}(p,q) \ - \ \tau(E^{(l)}(p,:))\right]_+
\end{equation}

\begin{equation}
\begin{gathered}
\tau(\bold{n}) \ = \ \frac{\left(\sum_{j\in Q(\bold{n})}\bold{n}_j\right) - 1}{| Q(\bold{n})|}\\
where \ \ Q(\bold{n}) \ = \ \{j \in [K] \ | \  \bold{n}_j > 0\}\\
\end{gathered}
\end{equation}
where $[K] = \{1, ..., K\}$ and $[t]_+ := max\{0, t\}$. All the coordinates that below threshold function $\tau(\cdot)$ will be truncated to zero. Thus, this piecewise function maintains the sparsity of adjacency matrix $A^{(l+1)}(p,q)$ . Moreover, as shown in figure \ref{fig:bio}, this link prediction is able to detect some underlying connections that are not given in the initialization step, which makes the model more robust.

\begin{figure}[h!]
\centering
\includegraphics[width=.5\textwidth]{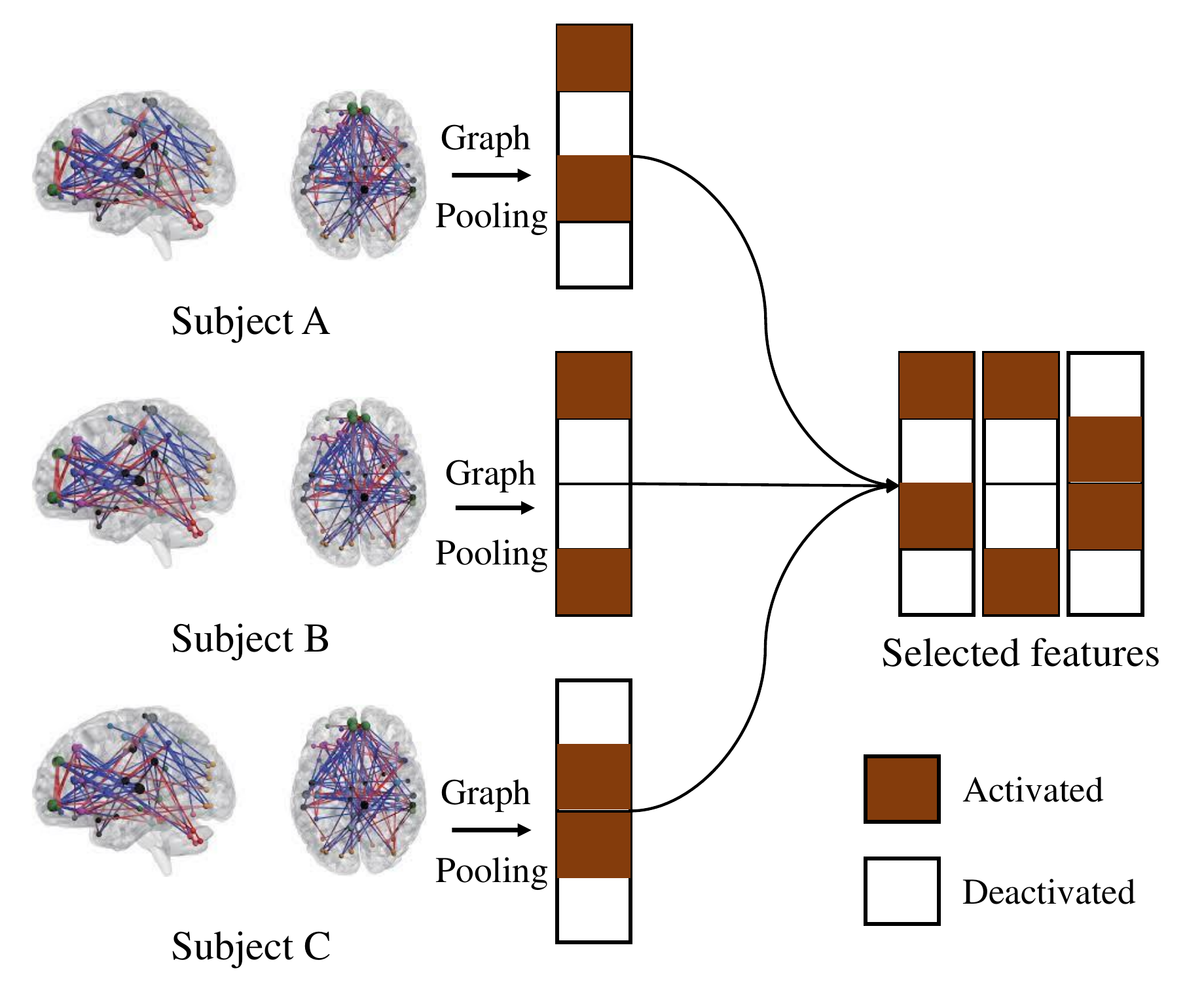}
\caption{Illustration of sparse feature expression. This step downsampled the brain imaging in terms of storage overhead but maintains the attribute of regions. Unlike previous global region selection methods, this approach retains the characteristics of subjects.}
\label{fig:sfe}
\end{figure} 

\subsubsection{Personalized Feature Extraction}\label{sec:pfe}
To conclude a universal ASD clinical diagnosis suggestion, \cite{parisot2018disease, wang2020aimafe, liu2020improved, mostafa2019diagnosis} tried to manually select top-N critical functional connections from the fMRI of all subjects. Ideally, this procedure would return some functional connections from which we could quickly tell if someone is suffering from this brain disorder. However, as discussed in \cite{parisot2018disease}, the results are not satisfactory: The highest mean accuracy is 70.40\%, and the smallest number of selected functional connections is 250, which is not adequate nor efficient for clinical diagnosis.

Different from that ambitious universal key features selection, we develop a personalized feature extraction strategy. Instead of selecting a set of universal biomarkers of ASD, we treat each subject separately and downsample the graph modeling according to its characters. As illustrated in the figure \ref{fig:sfe}, we performed graph pooling onto every individual and stored the results as sparse vectors. In terms of storage and computation cost, like the above hard selection, that strategy has successfully downsampled the input brain imaging. On the other hand, the feature matrix has preserved more information than the previous methods. The two main benefits of this sparse feature fusion are as follows: First, it downsamples the brain imaging to extraordinarily few key components even without a performance decrease, which will be discussed in the section \ref{sec:ipr}. Second, the difference in selected features between individuals has clinical meanings. It may indicate the varied brain regional activities and connections among different groups and will be discussed in the section \ref{sec:biomarker}.

\subsection{Graph Convolutional Networks}\label{sec:gcn}
To fuse brain imaging data and non-imaging data, we constructed a population graph where nodes represent subjects and edges indicate the similarity degree regarding phenotypic information. The non-imaging information similarity among subjects is characterized as the connectivity degree among nodes, i.e., nodes with similar phenotypic properties are more likely to be in the same community. We employ Graph Convolutional Networks to process the population graph structure with every node associated with a feature vector extracted from brain imaging. Proposed by \cite{kipf2016semi}, GCN extends convolution operations onto graph structures and is able to learn the node embeddings. At each layer of GCN, the node feature vector is then recalculated as the weighted sum of its and its neighbors' features, that is, the node embedding. Hence, the features of nodes that are in the same community tend to follow a similar distribution. Given the prior phenotypic information, this method has regularized the classification performance as shown in figure \ref{fig:ratio}.

\subsubsection{Population Graph Construction}
As stated in section \ref{subsection: abide}, we use the data of 871 subjects preprocessed by CPAC. The connection between two nodes is decided by their phenotypic similarity, i.e., gender, age, handedness, etc. However, caused by inconsistent measurement among different data collection sites, some categories of data are not or partly collected. For example, nearly $30\%$ of the handedness data are not available as illustrated in the table \ref{table:data}. Suggested by \cite{parisot2018disease}, we consider a subset of the whole non-imaging data, which contains gender, age, and data collection sites information. Intuitively, the similarity is computed as the cosine similarity between two phenotypic feature vectors $M_u$ and $M_v$.

\begin{equation}
Sim(u, v) \ = \ \lvert \frac{M_u \cdot M_v}{\left\|M_u\right\|\left\|M_v\right\|} \rvert
\end{equation}
where $M = \{Age, Gender, Site\}$ denotes the selected subset of non-imaging data. A threshold of 0.5 is then applied to the derived similarity values to decide whether the two nodes $u, v$ are connected or not.
\begin{equation}
A(u, v) \ = \ \left\{
\begin{aligned}
1, & & if \ {Sim(u, v)>0.5}\\
0, & & otherwise \\
\end{aligned}
\right.
\end{equation}
where $Sim(u, v)$ is the similarity score of the two subjects. $A$ represents the adjacency matrix of the graph. Two nodes are connected if their cosine similarity value is above 0.5. By these means, the population graph is initialized as an undirected graph containing 871 nodes.

\subsubsection{GCNs}
We have implemented two kinds of GCN in this part. The first layer is the same as the one proposed by \cite{kipf2016semi}. The second layer is the Cluster-GCN presented by \cite{chiang2019cluster}, which has accelerated the basic GCN block.

\begin{table*}[t!]
\caption{Comparison with state-of-the-art methods on ABIDE I dataset}
\centering
\label{table:result}
\begin{adjustbox}{center}
\begin{threeparttable}

\begin{tabular*}{.8\textwidth}{@{\extracolsep{\fill}}lcccc}
\toprule
\multirow{2}{*}{References} & \multirow{2}{*}{Methods} &  \multicolumn{3}{c}{Performance} \\
\cmidrule{3-5}
&&Accuracy & Sensitivity & Specificity \\
\midrule
\cite{parisot2017spectral} & GCN & 69.50 & - & - \\
\cite{dvornek2017identifying} & LSTM & 66.80  & - & - \\
\cite{abraham2017deriving} & SVC & 66.80  & 61.00 & 72.30 \\
\cite{heinsfeld2018identification} & DNN & 70.00  & 74.00 & 63.00 \\
\cite{parisot2018disease} & MLP and GCN & 70.40  & - & - \\
\cite{khosla20183d} & 3D-CNN & 73.30  & - & - \\
\cite{eslami2019asd} & Autoencoder & 67.50  & 68.30 & 72.20 \\
\cite{mostafa2019diagnosis} & FCs selection and LDA & 77.70  & - & - \\
\cite{jiang2020hi} & Joint learning & 73.10  & 71.40 & 74.60 \\
\cite{sherkatghanad2020automated} & CNN & 70.20  & 77.00 & 61.00 \\
\cite{liu2020improved} & FCs selection and SVM & 76.80  & 72.50 & 79.90 \\
\cite{wang2020aimafe} & FCs selection and MLP & 74.52  & 80.69 & 66.71\\
\cmidrule{1-5}
\multirow{2}{*}{\textbf{Present study}} &  Graph pooling and LR & \textbf{87.62} & \textbf{86.76} & 88.36\\
 &   Graph pooling and GCN & 86.07  & 82.88 & \textbf{88.80}\\
\bottomrule
\end{tabular*}

\begin{tablenotes}
\footnotesize
\item[] GCN: Graph Convolutional Networks
\item[] LSTM: Long Short-Term Memory
\item[] SVC: Support Vector Classification
\item[] MLP: Multi-Layer Perceptron
\item[] LDA: Linear Discriminant Analysis
\item[] LR: Logistic Regression
\item[] FCs: Functional Connections
\end{tablenotes}

\end{threeparttable}
\end{adjustbox}
\end{table*} 

Extending convolution operations to non-Euclidean space, GCNs have achieved promising performance on arbitrarily structured graphs. Though there exist different forms of GCN block, the universal core task is to learn a non-linear function $f(H^{(l)}, A)$ which aggregates the feature vectors of connected nodes to generate features for next layer:
\begin{equation}
H^{(l+1)} \ = f(H^{(l)}, A)  \ =  \ \sigma(\tilde{D}^{-\frac{1}{2}}\tilde{A}\tilde{D}^{-\frac{1}{2}}H^{(l)}W^{(l)})
\end{equation}
with $\tilde{A}=A+I$, where $I$ is the identity matrix and $\tilde{D}$ is the diagonal node degree matrix of $\tilde{A}$. For the $l$-th layer of the GCN, the graph can be represented as the feature matrix $H^{(l)}$. $H^{(0)}=X$ and $H^{(L)}=Z$ denote the input and final output feature matrix respectively. $W^{(l)}$ is the learnable weight matrix and $\sigma(\cdot)$ is the non-linear activation function, ReLU. In this way, the features are aggregated to form features of the next layer. \cite{chiang2019cluster} reduced the computational cost by clustering nodes into multiple bathes:

\begin{equation}
\frac{1}{|\mathcal{B}|}\sum_{i\in\mathcal{B}}{\nabla loss(y_i, z_i^L)}
\end{equation}
where $\mathcal{B}$ indicates the subset of nodes and $z_i^L$ represents the final prediction label of the $i$-th subject. Hence, at the loss back-propagation phase, the model only needs to calculate the gradient for the mini-bath. The Binary Cross-Entropy Loss is defined as the loss function:

\begin{equation}
loss(y_i, z_i^L) \ = \ -[y_i * log(z_i^L) + (1-y_i)*log(1-z_i^L)]
\end{equation}

After the graph convolutional layers, a linear classifier is applied on each node. The final outputs of the classifier represent the probability of ASD. By filtering the probabilities with a 0.5 threshold, the model finally outputs the predicted diagnostic group of each subject.

\section{Experimental analysis}
\subsection{Experimental Settings}
To make this experiment consistent with other studies \cite{khosla20183d, mostafa2019diagnosis, sherkatghanad2020automated, parisot2018disease, parisot2017spectral, dvornek2017identifying}, we have performed the 10-fold cross-validation on the 871 samples and repeated it ten times. The multilayer perceptron and graph convolutional networks are trained separately but strictly on the same train set. At the training stage of the multilayer perceptron, we have employed the nested 10-fold cross-validation and repeated the inner loop five times every outer one. The whole framework is trained and tested on an NVIDIA TESLA V100S. During the optimization, we have utilized the Adam optimizer, of which the parameters are set as follows: learning rate = 0.0001, weight decay = 0.01. We have also used the dropout to enhance the generalization of GCN with a dropout ratio of 0.01.

\subsection{Results}
In the table \ref{table:result}, we have compared our framework with other models on the same ABIDE I dataset preprocessed by CPAC \cite{craddock2013neuro}. In general, those methods can be categorized into two types: single-stage and multi-stage. The single-stage methods directly deploy deep learning methods, like CNN, to deal with this ASD VS Control binary image classification problem\cite{dvornek2017identifying, abraham2017deriving, heinsfeld2018identification, khosla20183d, eslami2019asd, sherkatghanad2020automated}. However, limited by the number of available training samples, these naive implementations of neural networks have not achieved promising performance. On the other hand, multi-stage methods usually consist of two components: feature extraction and classification. Previous works trained feature extractors to extract features from brain functional connections\cite{parisot2018disease, mostafa2019diagnosis, liu2020improved, wang2020aimafe}. A classifier, like SVC, is then trained with extracted features as inputs. This kind of framework has successfully downsampled the high-dimensional brain imaging and thus made obvious performance improvement even with linear classifiers compared to the straightforward CNN models. For example, \cite{mostafa2019diagnosis} constructed a specific workflow to select the key features from brain imaging and achieved an accuracy of 77.7$\%$ with linear discriminant analysis.

\begin{figure}[htbp!]
\centering
\includegraphics[width=.5\textwidth]{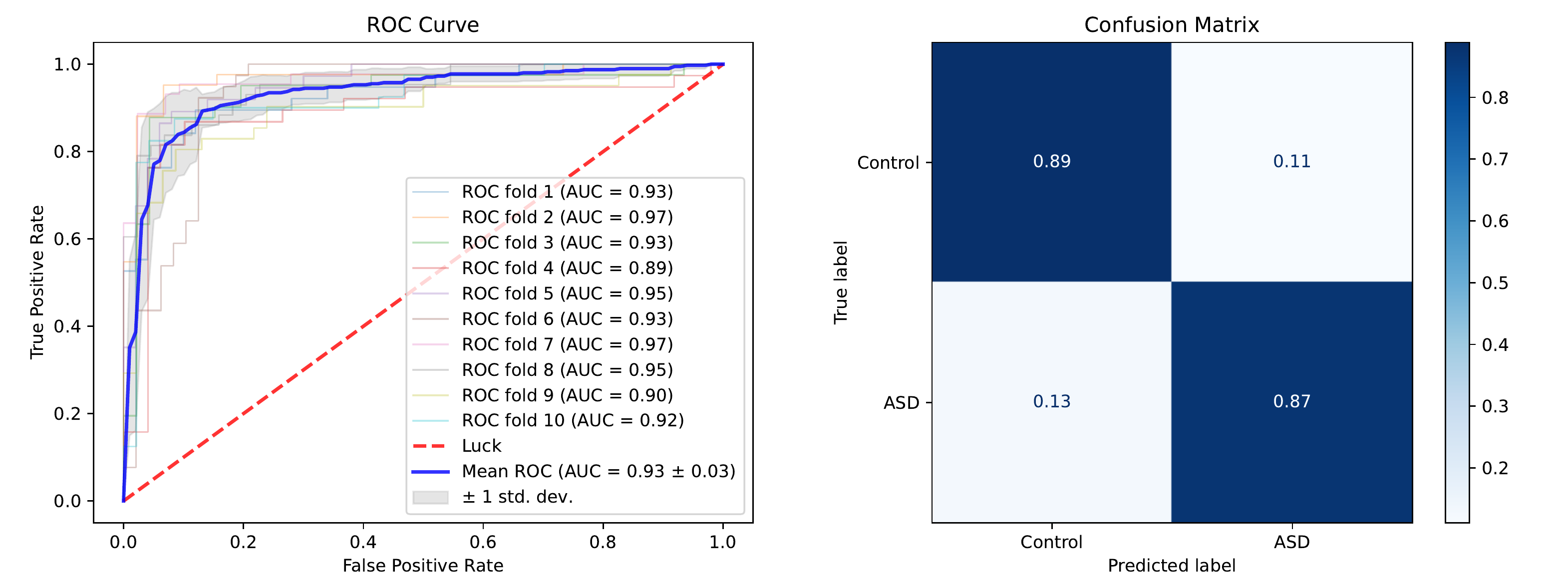}
\caption{Example of framework performance on the outer loop of the nested cross-validation. The backbone is set as Graph pooling and LR with pooling ratio = 0.01 and the random seed of outer loop = 168.}
\label{fig:result}
\end{figure}

\begin{figure*}[t!]
\centering
\includegraphics[width=\textwidth]{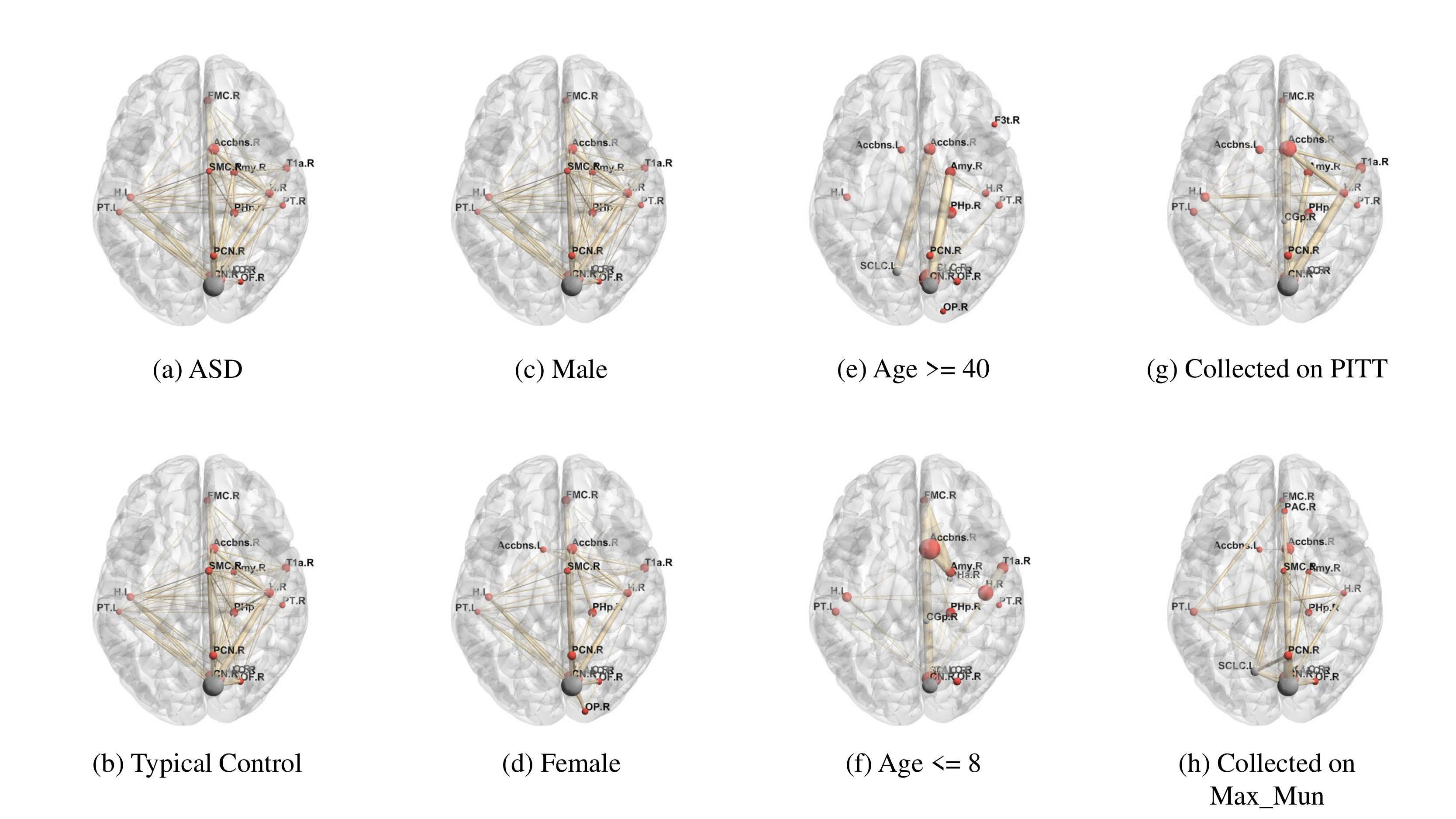}
\caption{Result of graph pooling for different groups. The pooling ratio is set as 0.05, i.e., 6 regions are selected out of 111. Each axial view of the brain shows the top 15 nodes most frequently selected from those who are inside the corresponding group. The width of edges and size of nodes indicate the relative frequency of being selected. Number of subjects inside each group are as follows: a: 403; b: 468; c: 727; d: 144; e: 14; f: 15; g: 50; h: 46.}
\label{fig:bio}
\end{figure*}

Like the above multi-stage methods, the present framework includes feature extraction and classification parts. We have employed graph pooling to downsample the given brain networking and trained a multilayer perceptron to further extract features, whereas previous feature extractors can only extract features from functional connections. Inspired by \cite{parisot2018disease}, we employ GCN in the final classification part, which has leveraged imaging and non-imaging information. By these means, our framework outperforms the state-of-the-art method, reaching an accuracy of 87.62$\%$ and a mean AUC of $0.92$. The efficiency of graph pooling is discussed in the following part.

\subsection{Efficiency of Graph Pooling}\label{sec:ipr}
	As previously mentioned, studies have reported abnormal brain functional connections found on ASD subjects\cite{villalobos2005reduced}. With this prior knowledge, in \cite{parisot2018disease, mostafa2019diagnosis, liu2020improved, wang2020aimafe}, the authors solely simplified brain imaging as functional connections. Specifically, they computed correlation coefficients between every two regions and fed them into feature extractors. This correlation matrix representation of the brain successfully downsampled raw fMRI data, resulting in crucial information loss as the regional features were omitted. In contrast, graph pooling extracts features directly from the graph representation of the brain. As shown in the table \ref{table:gp}, graph pooling has outperformed the state-of-the-state method, reaching an accuracy of $87.62\%$ even with linear regression as a classifier. Especially in \cite{abraham2017deriving}, the authors have developed a similar workflow and achieved an accuracy of $66.80\%$. The only difference is that they extracted features from functional connections, whereas graph pooling does it by downsampling the entire brain imaging.

\begin{table}[h!]
\caption{The accuracy of ABIDE I classification with linear classifiers for different feature extraction methods}
\centering
\label{table:gp}
\begin{adjustbox}{center}
\begin{threeparttable}

\begin{tabular*}{.5\textwidth}{@{\extracolsep{\fill}}lcc}
\toprule
References & Feature extractor and classifier &  Accuracy \\
\midrule
\cite{nielsen2013multisite} & FCs and SVC & 60.00\\
\cite{abraham2017deriving} & MLP and SVC & 66.80\\
\cite{wang2020aimafe} & FCs selection and MLP & 74.52\\
\cite{liu2020improved} & FCs selection and SVM & 76.80 \\
\midrule
\textbf{Present study} & \textbf{Graph pooling and LR} & \textbf{87.62}\\
\bottomrule
\end{tabular*}

\end{threeparttable}
\end{adjustbox}
\end{table}

\begin{figure*}[htbp!]
\centering
\subfigure[Accuracy]{\includegraphics[width=.32\textwidth]{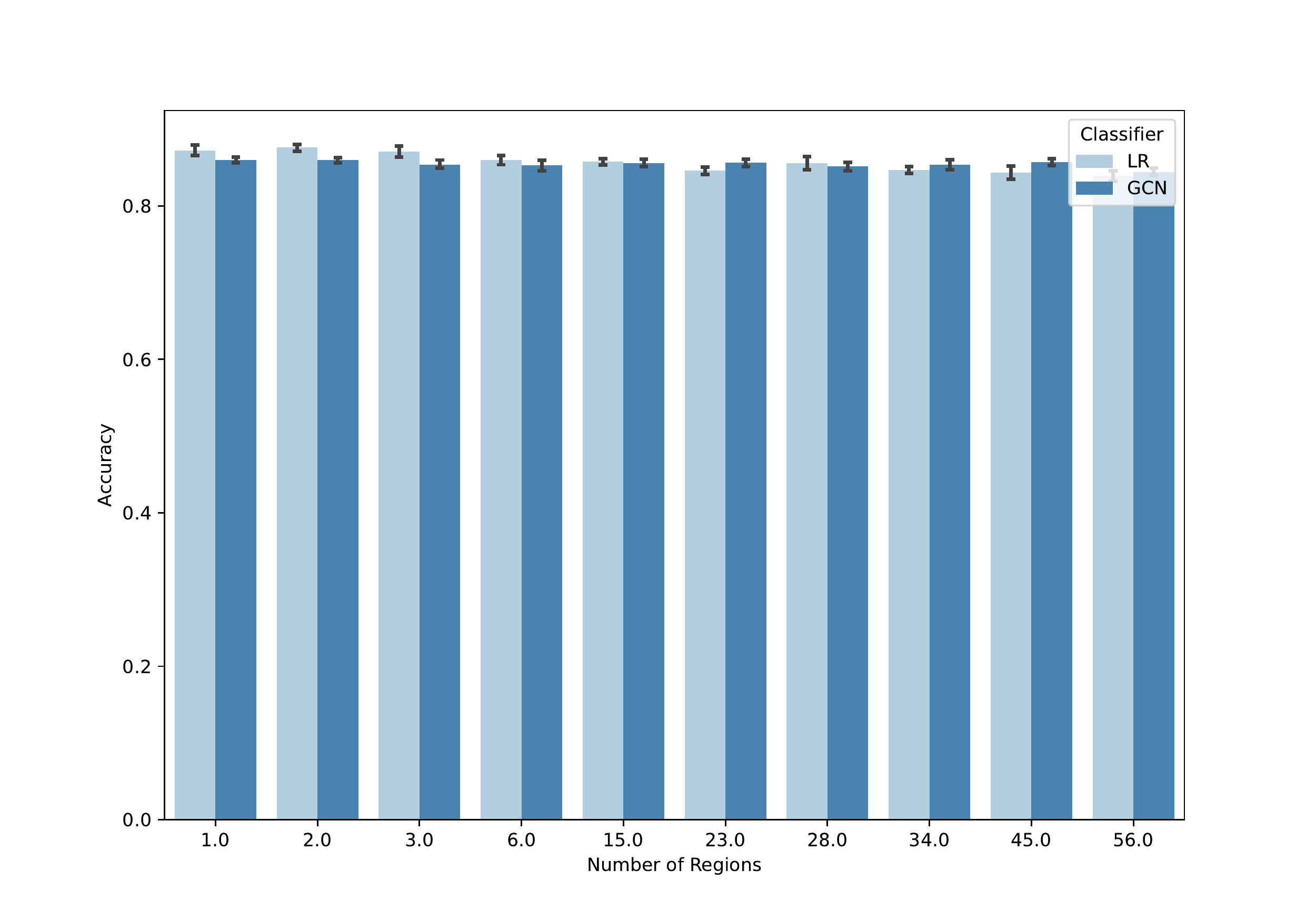}}
\subfigure[Sensitivity]{\includegraphics[width=.32\textwidth]{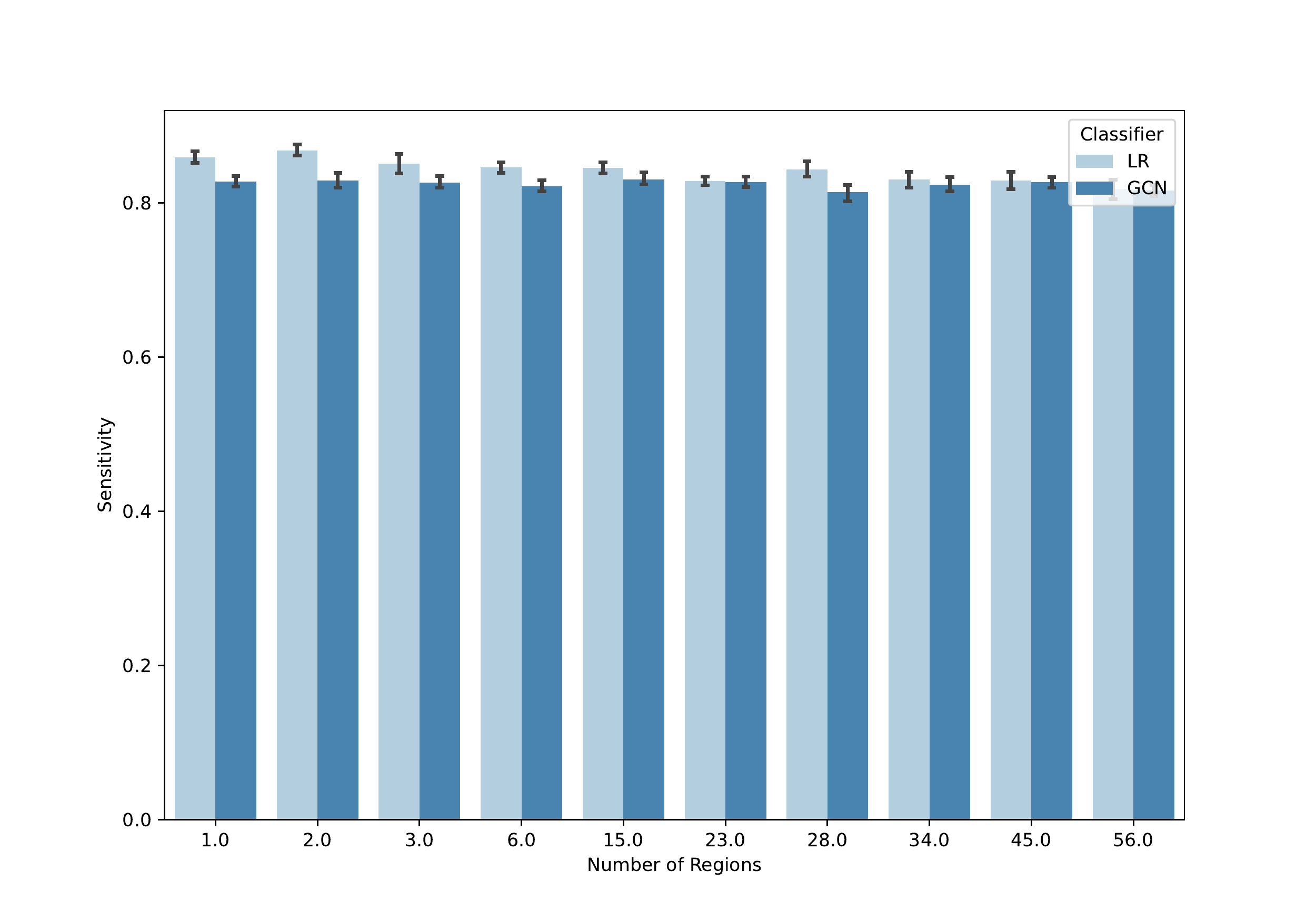}}
\subfigure[Specificity]{\includegraphics[width=.32\textwidth]{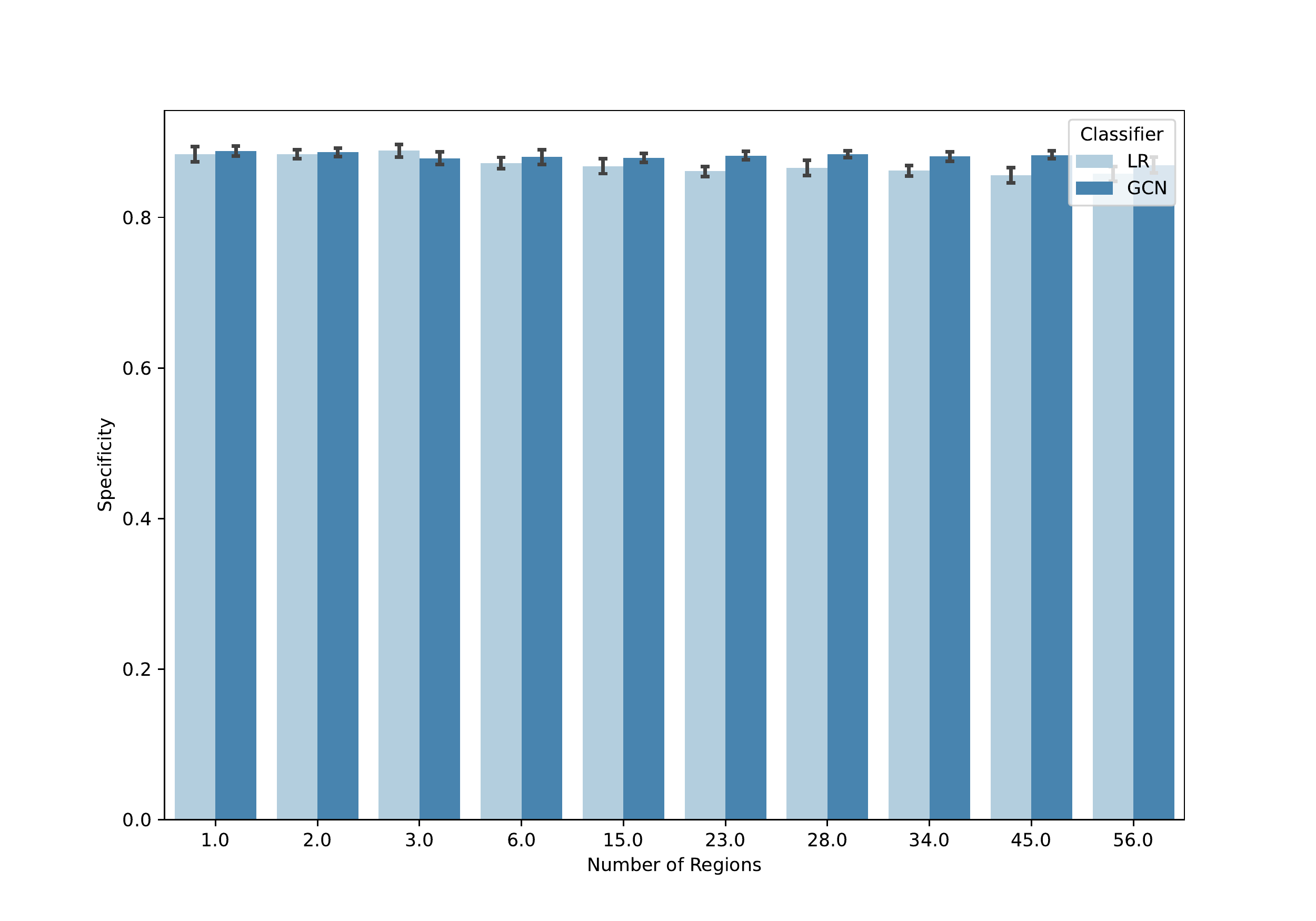}}
\caption{ABIDE I classification performance for the different numbers of brain regions selected at the graph pooling stage. Both the Logistic Regression and Graph Convolutional Networks work under the same settings.}
\label{fig:ratio}
\end{figure*}

To further evaluate the efficiency of graph pooling, we have tested the graph pooling progress for different values of the pooling ratio. Previous functional connection selection methods chose a certain number of edges from the graph representation of the brain\cite{parisot2018disease, mostafa2019diagnosis}. That inflexible strategy disregarded the heterogeneity of brain activities caused by inter-individual differences and even occasional personal status. Thus, that static procedure failed to achieve a satisfactory classification performance. As previously discussed in section \ref{sec:pfe}, the self-attention mechanism inside graph pooling and sparse feature extraction have realized personalized brain essential components selections. Illustrated in figure \ref{fig:ratio}, the classification performance of the framework is not sensitive to the pooling ratio when GCN is enabled to regularize the sample distributions. Even having downsampled the brain imaging to one region, the present model can still reach a classification accuracy of 85.98\%.

The main advantages of graph pooling are 3-fold: First, it can be easily generalized to other related problems. Previous methods work under a strong assumption that it is only the abnormal functional connections that cause ASD. This presupposition is not only biased in the current task \cite{saenz2020adhd,harris2006brain, hadjikhani2007abnormal, kim2015abnormal} but also limit generalizing these methods to other brain disorder diagnosing problems. On the contrary, graph pooling requires less prior knowledge about brain functions as it straight receives the entire graph representation of brain imaging and can learn connections beyond the given brain networking. Besides, as discussed in section \ref{sec:hgp}, this unsupervised downsampling method needs no training overhead. Second, it is more efficient for brain feature extraction. As shown in table \ref{table:result}, using features selected by graph pooling, a linear classifier can achieve much better performance than other more complex models did. In figure \ref{fig:ratio}, we illustrate that the extracted features are still reliable even with the whole brain imaging downsampled to several connected nodes. Third, it can be aware of individual characters to some degree, which is benefited from the self-attention mechanism and sparse representation of extracted features, as discussed in section \ref{sec:pfe}. We have observed variance in brain imaging pooling results of subjects from different groups, as shown in figure \ref{fig:bio}, and found even more obvious variance when we further split the groups. This inter-group heterogeneity may have clinical and biomedical meanings. 

\begin{figure}[htbp!]
\centering
\subfigure[Logistic Regression]{\includegraphics[width=.45\textwidth]{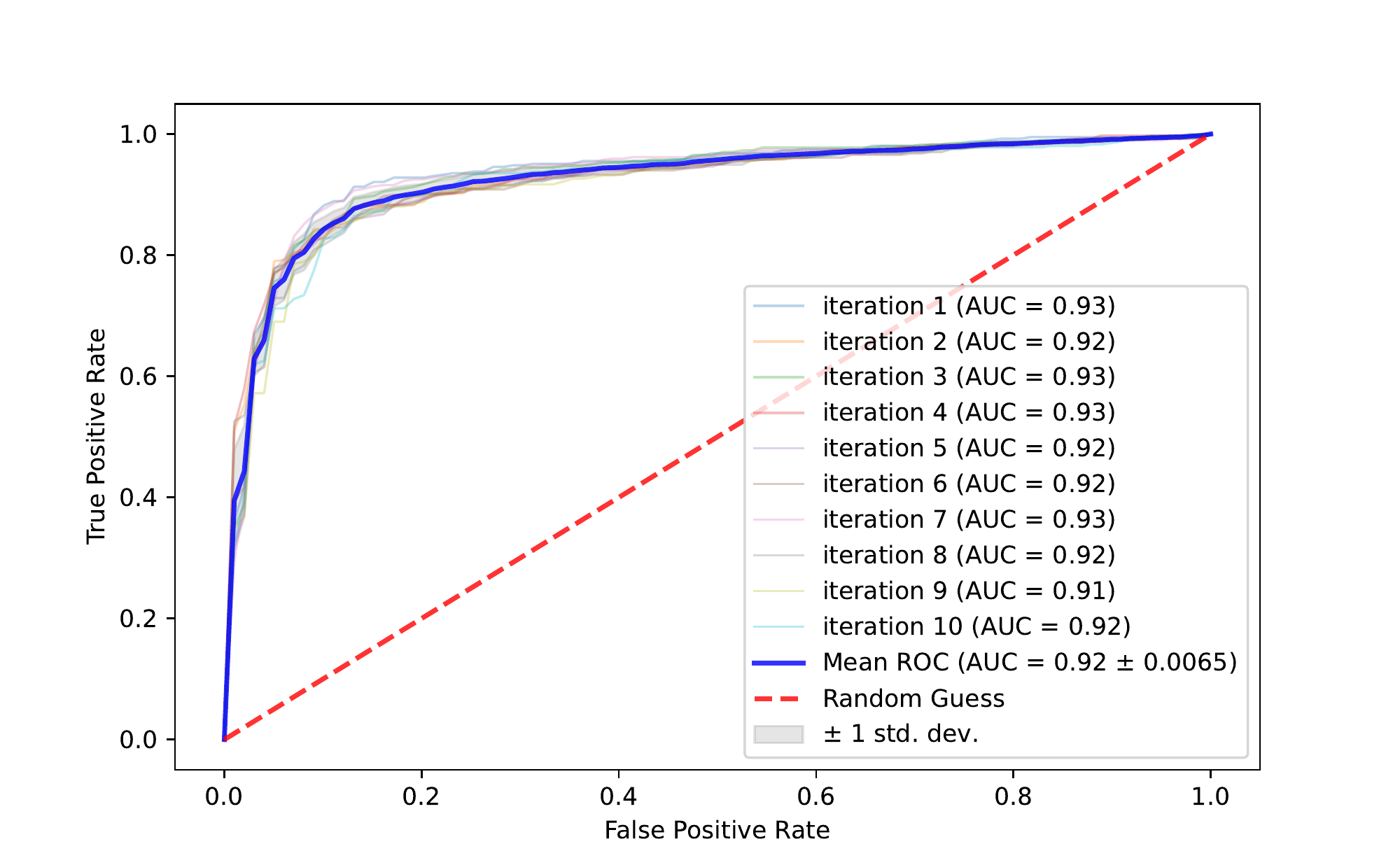}}\\
\subfigure[Graph Convolutional Networks]{\includegraphics[width=.45\textwidth]{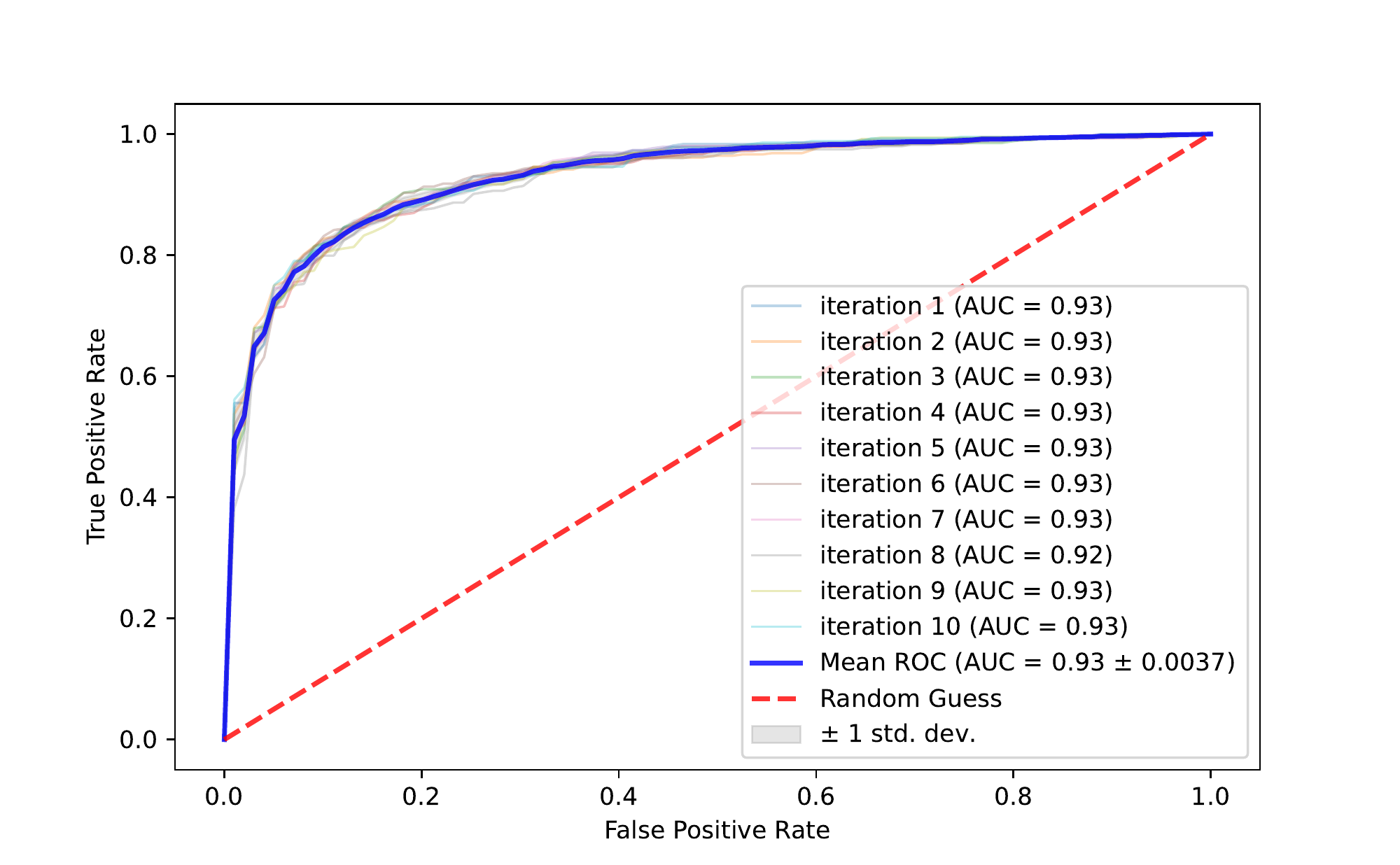}}
\caption{Area Under the ROC Curve of the two classifiers. The out loop of the nested cross-validation is repeated 10 times. For each iteration, the AUC is measured as the mean value of the 10 folds.}
\label{fig:lrVSgcn}
\end{figure}

\subsection{Key Brain Substructures}
Interpreting the ASD diagnosis framework has been being a hot topic as it may indicate the brain biomarkers of autism spectrum disorder and direct the early intervention. To intuitively exhibit the pooling results, i.e., the most critical substructures selected at that stage, we have plotted them by BrainNet Viewer proposed by \cite{xia2013brainnet}. We have applied graph pooling onto the brain imaging of all subjects with a pooling ratio of 0.05. Thus, for every individual, the pooling result is six selected nodes and their connections. Subsequently, we have split 871 subjects into different groups regarding their non-imaging properties, including age, gender, and the data collection site they belong to. We have surveyed the top 15 most frequently selected regions and connections from the pooling results of all subjects inside this group for each one.
\label{sec:biomarker}
\begin{figure}[h!]
\centering
\includegraphics[width=.5\textwidth]{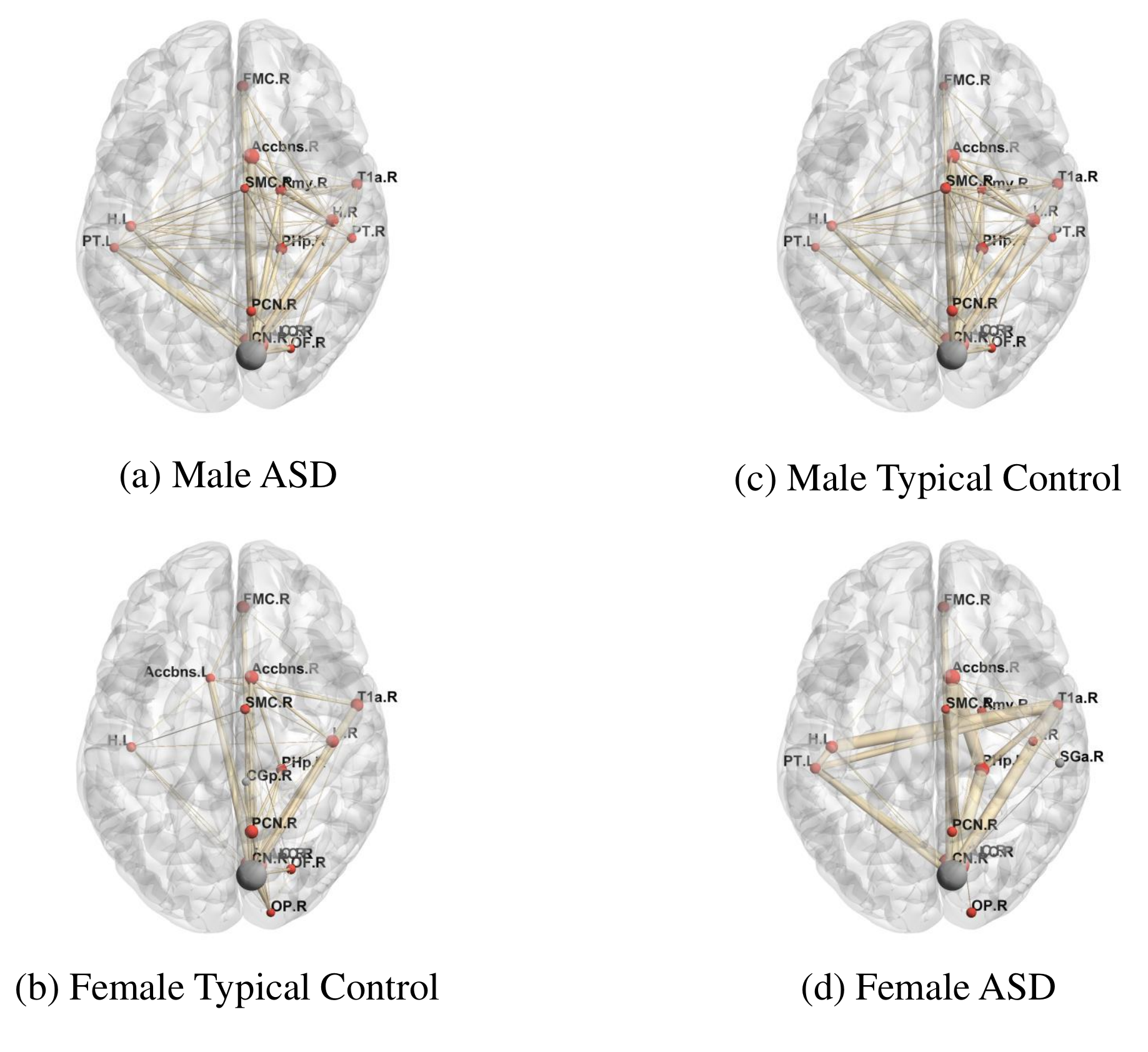}
\caption{Additional brain view for figure \ref{fig:bio} with same settings. The illustrated subgraphs indicate selection preference in graph pooling.}
\label{fig:add}
\end{figure}

\begin{figure*}[htbp!]
\centering
\subfigure[Logistic Regression]{\includegraphics[width=.48\textwidth]{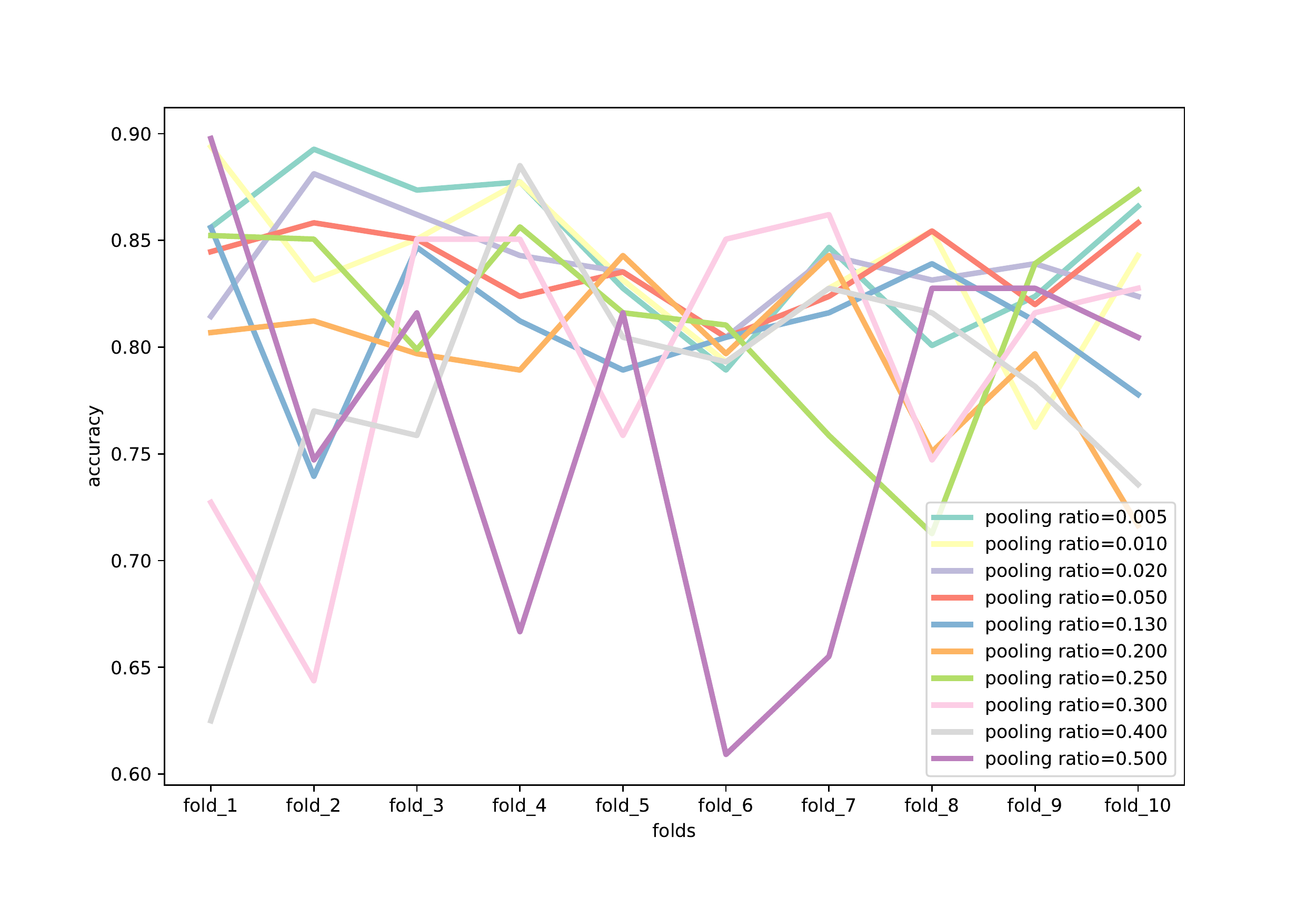}}
\subfigure[Graph Convolutional Networks]{\includegraphics[width=.48\textwidth]{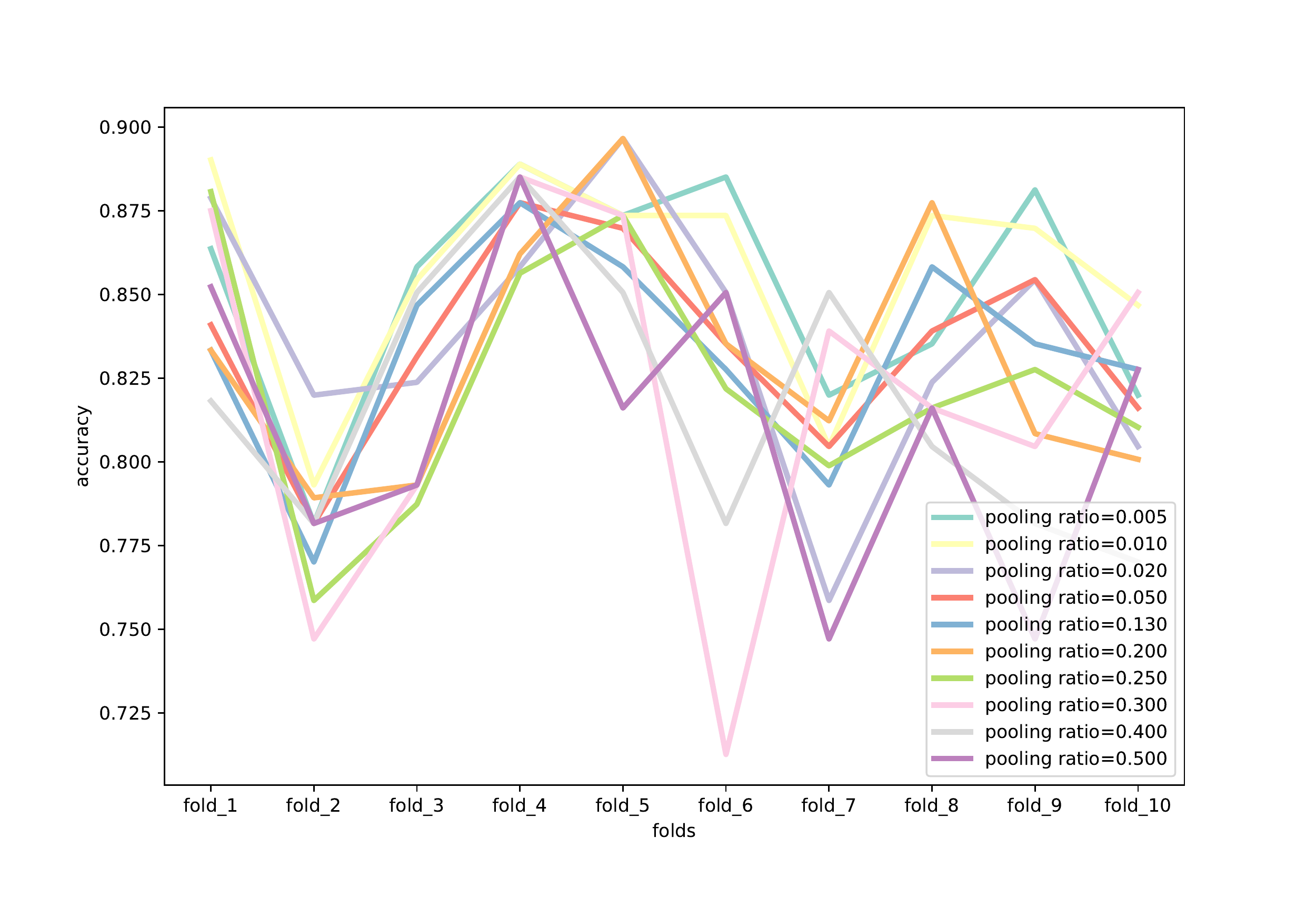}}
\caption{Impact of different training and testing sets for Logistic Regression and Graph Convolutional Networks. The two backbones are evaluated under different pooling ratios of graph pooling. For each pooling ratio, the 10-fold cross-validation is repeated 10 times with the same dataset split (random seed = 13).}
\label{fig:folds}
\end{figure*}

Unlike figuring out universal brain biomarkers of ASD, the outputs of self-attention pooling only have specified the importance of regions and edges of individual brain imaging. However, we can still draw some conclusions by summarizing the pooling results in different groups. The illustrated substructures of the brain, as shown in figure \ref{fig:bio}, may indicate a common brain activity mechanism inside a specific group.

Ideally, as discussed in section \ref{sec:hgp}, the remained regions are the first activated ones regarding external stress or active internal activities. They act like a trigger that has launched a sequence of regional activities. Based on this knowledge about model working principles, we have observed some inter-group heterogeneity in essential substructures selected by graph pooling. That finding indicates that self-attention graph pooling along can be aware of individual phenotypic properties to some degree. In figure \ref{fig:bio}, few differences have been found between ASD and Typical Control subjects as the heterogeneity caused by individual characteristics may be averaged. We further divide the two groups into four by incorporating gender into consideration, as shown in figure \ref{fig:add}. According to the basic posist above, we may not conclude that it is the illustrated brain key substructures differences between different groups that have caused ASD. But these personalized subgraphs, serving as the inputs of MLP, are sufficient for ASD diagnosis.

\subsection{Efficiency of Graph Convolutional Networks}

In \cite{parisot2018disease}, the authors incorporated graph convolutional networks with brain functional selections and obtained accuracy improvement compared to \cite{abraham2017deriving}. The classification accuracy was increased from 66.80\% to 70.40\% by leveraging both imaging and non-imaging information with GCN. To figure out the efficiency of GCN in our framework, we have compared it with Logistic Regression under 10-fold cross-validation with the same dataset splits. As indicated in figure \ref{fig:ratio}, the two classifiers, GCN and LR, have reached the highest accuracy when the pooling ratio is set as 0.01. Herein, we have followed that observation and without changing other settings.

\begin{table}[h!]
\caption{Compare classifiers: Linear Regression and Graph Convolutional Networks. The training overhead was calculated by training the both on one fold out of 10-fold cross-validation.}
\centering
\label{table:lrVSgcn}
\begin{adjustbox}{center}
\begin{threeparttable}

\begin{tabular*}{.5\textwidth}{@{\extracolsep{\fill}}lcc}
\toprule
Measurement & Linear Regression &  GCN \\
\midrule
Mean Accuracy (\%) & \textbf{87.62} & 86.07\\
Standard Deviation of Accuracy & 0.055 & \textbf{0.027}\\
Area Under the ROC Curve & 0.92 & \textbf{0.93}\\
Standard Deviation of AUC & 0.0065 & \textbf{0.0037}\\
Average Training Overhead (s) & \textbf{0.25} & 1423.26\\
Average Training Epochs & \textbf{625} & 76169 \\
\bottomrule
\end{tabular*}	

\end{threeparttable}
\end{adjustbox}
\end{table}

\begin{figure}[htbp!]
\centering
\subfigure[Raw Extracted Features; Sites]{\includegraphics[width=.24\textwidth]{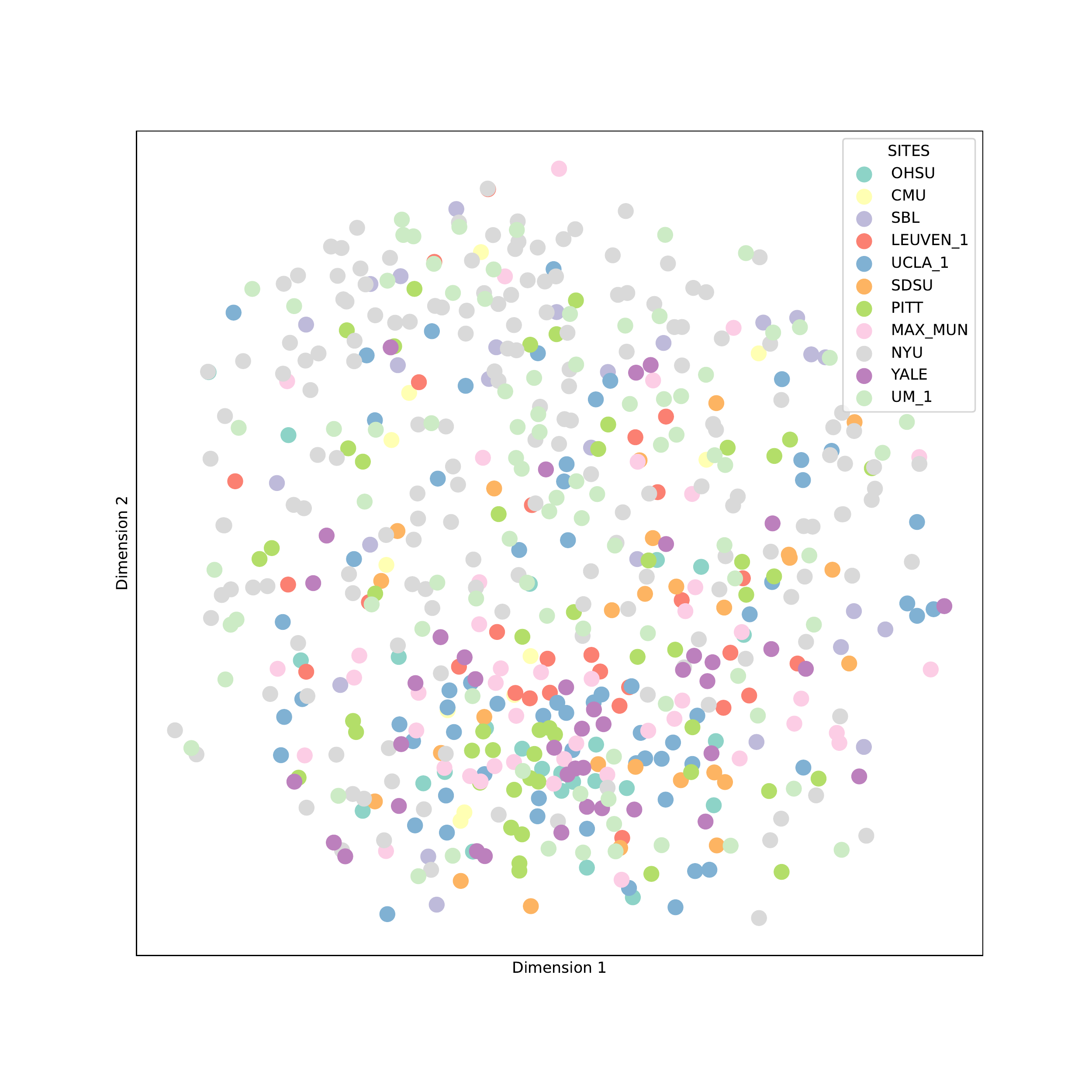}}
\subfigure[Node Embeddings; Sites]{\includegraphics[width=.24\textwidth]{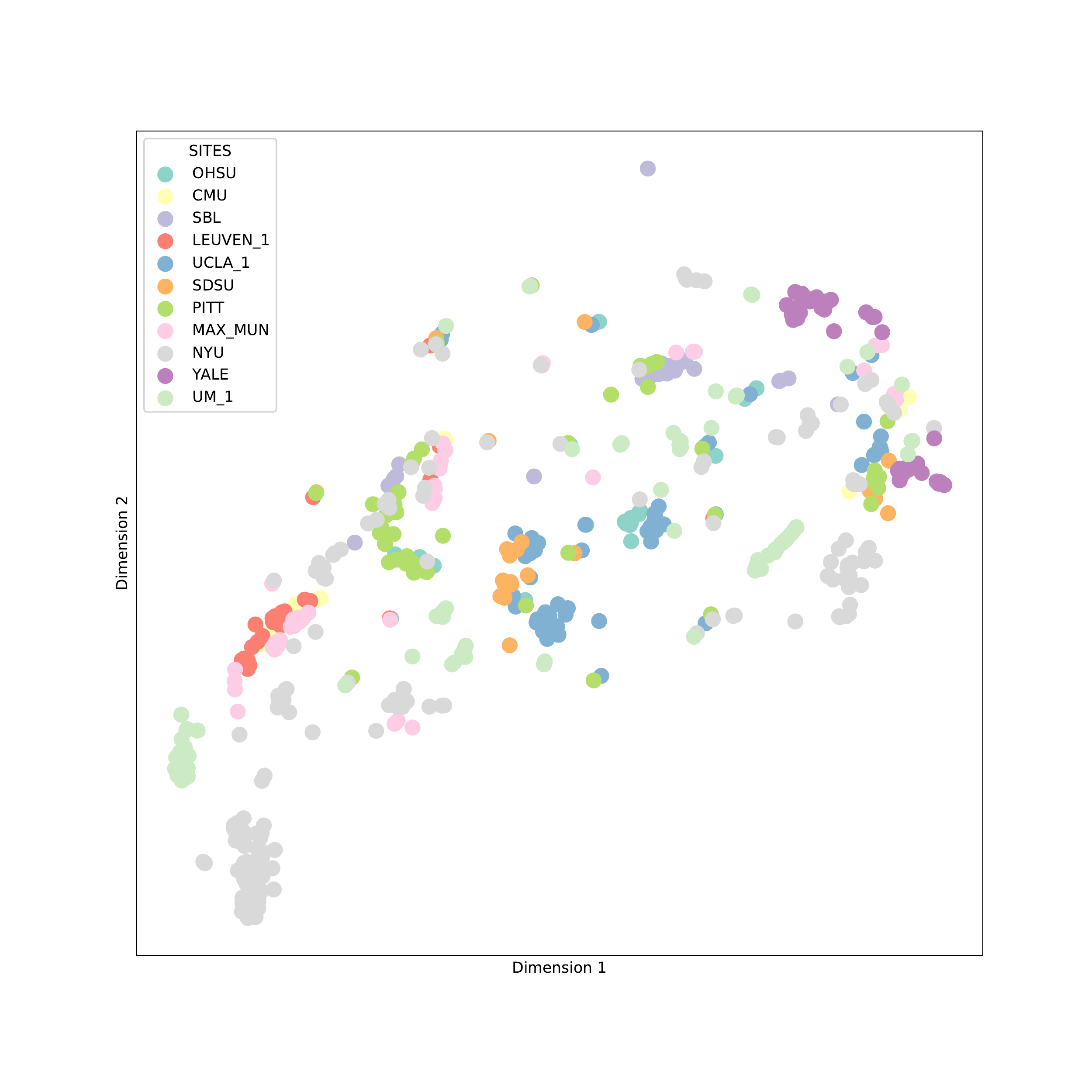}}
\\
\subfigure[Raw Extracted Features; Genders]{\includegraphics[width=.24\textwidth]{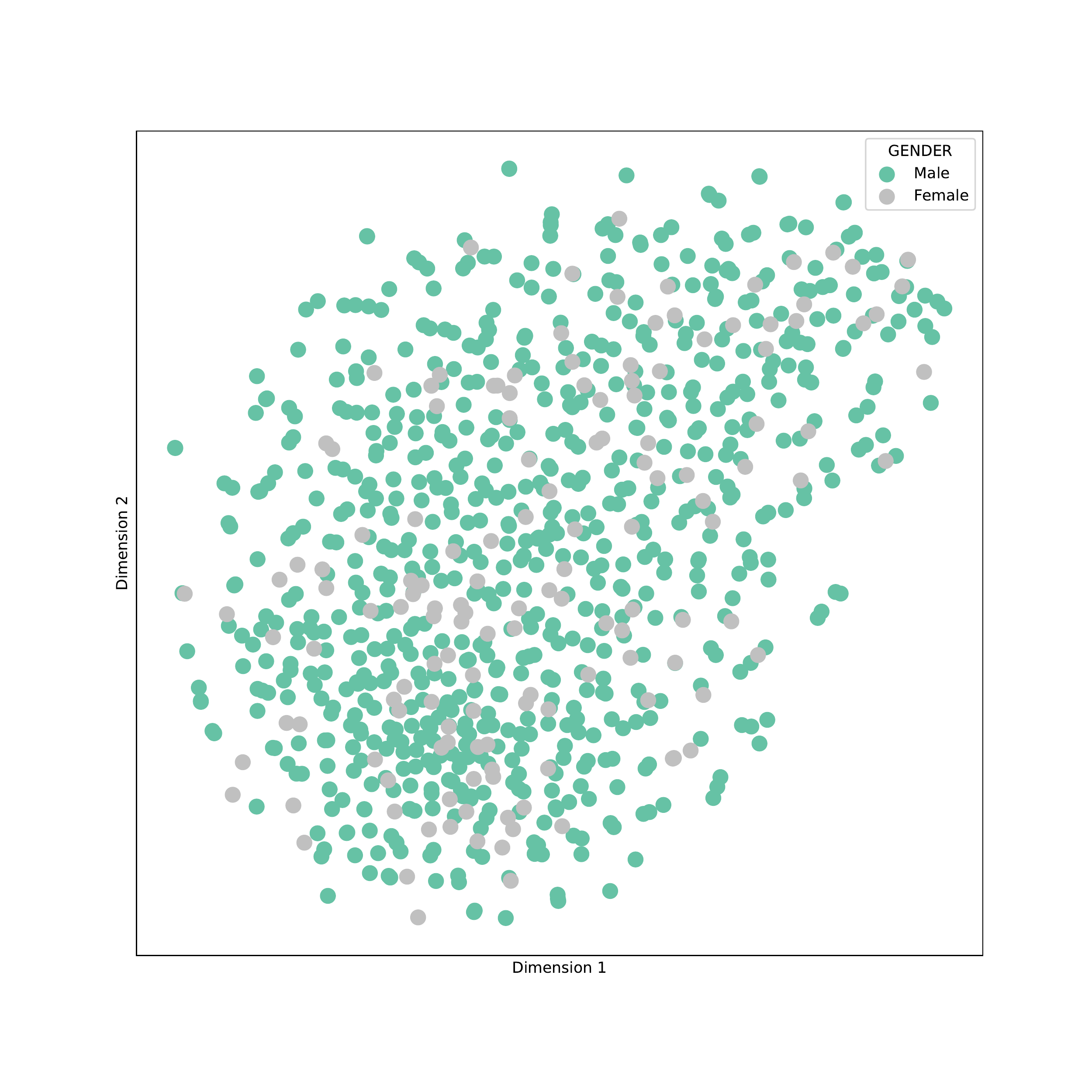}}
\subfigure[Node Embeddings; Genders]{\includegraphics[width=.24\textwidth]{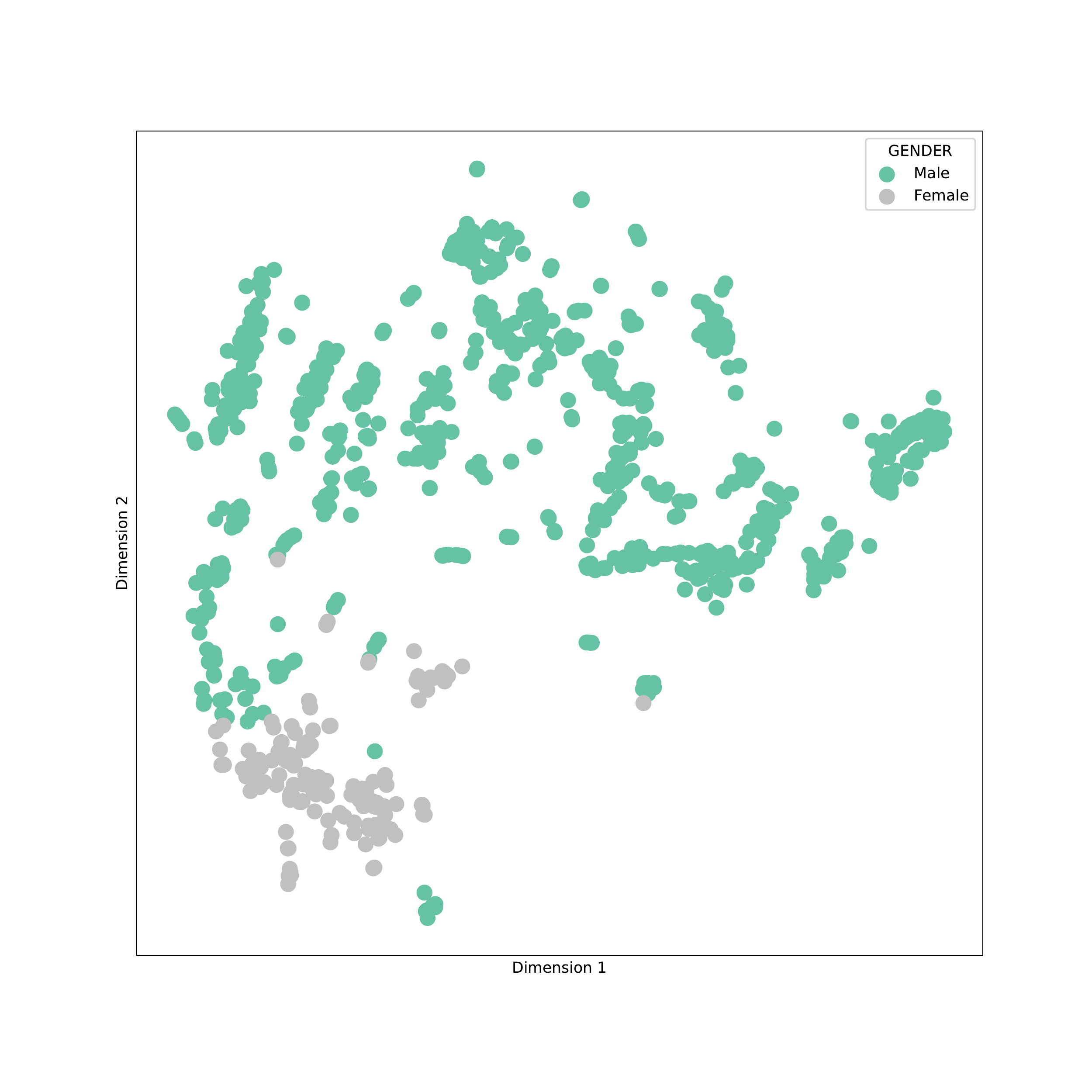}}
\\
\subfigure[Raw Extracted Features; Ages]{\includegraphics[width=.24\textwidth]{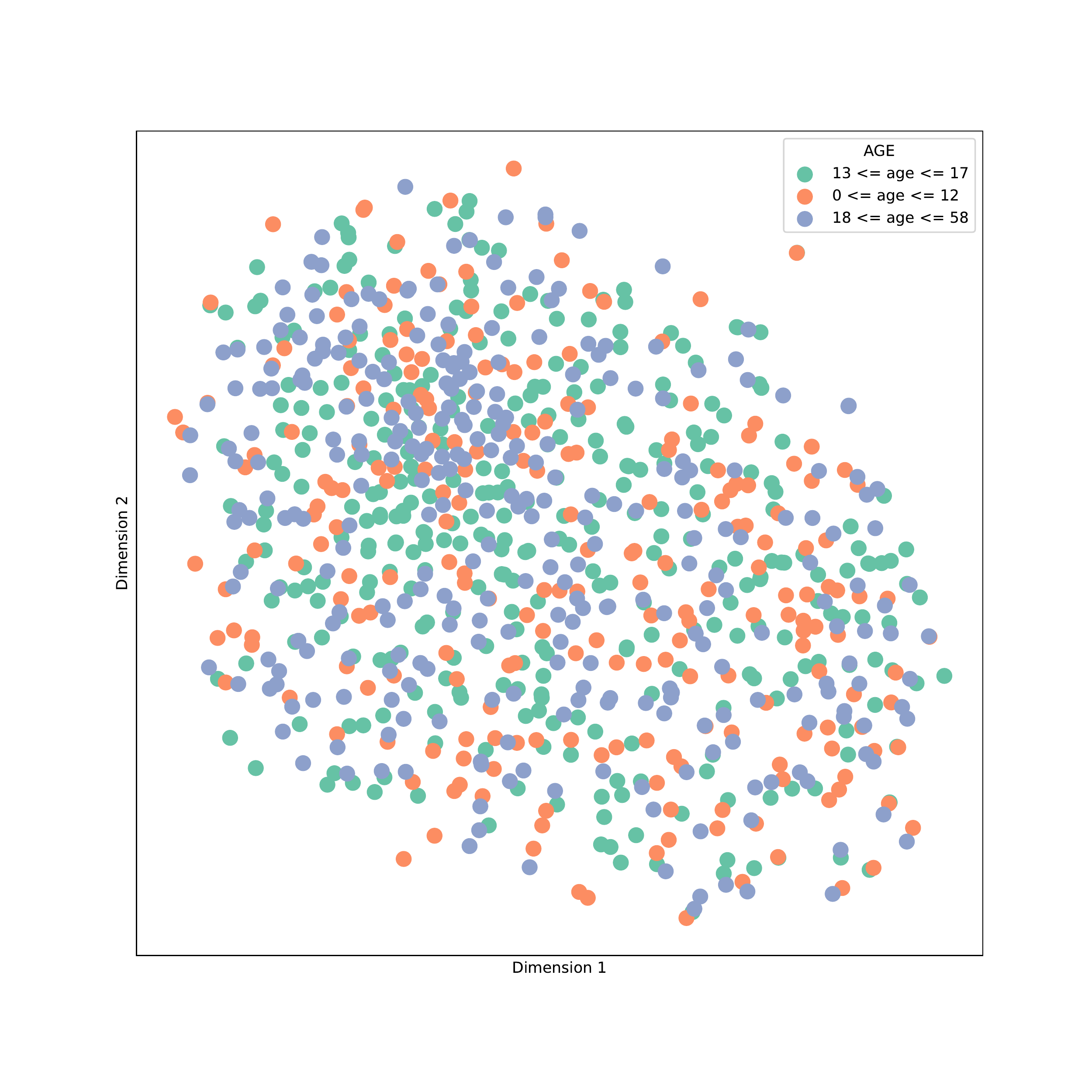}}
\subfigure[Node Embeddings; Ages]{\includegraphics[width=.24\textwidth]{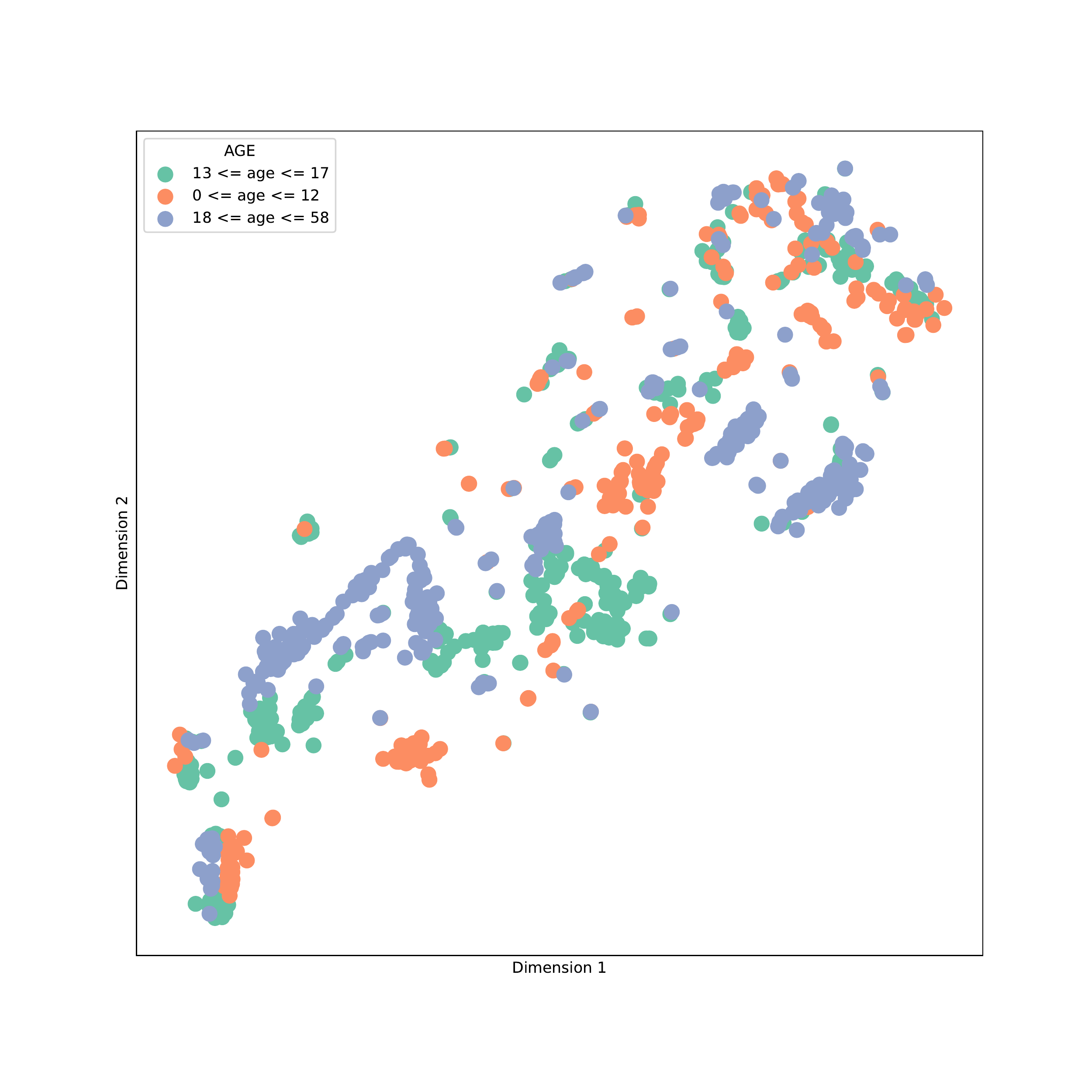}}
\caption{2D view of the node embeddings learned by Graph Convolutional Networks. The nodes, which represent subjects, are colored according to different phenotypic properties: Sites, genders, and ages}
\label{fig:2d}
\end{figure}

As illustrated in figure  \ref{fig:lrVSgcn} and figure \ref{fig:result}, compared to logistic regression, the classification performance of GCN is more robust in terms of accuracy, specificity, sensitivity, and AUC. This regularization impact can also be observed in figure \ref{fig:folds}, i.e., when the training and testing sets are fixed, the GCN is more likely to reproduce the same performance. This hypothesis is also supported by the superior of GCN concerning the standard deviation of both accuracy and AUC in table \ref{table:lrVSgcn}, which has been reported by \cite{parisot2018disease, parisot2017spectral} as well.

All the previous works, which have employed GCN for the same purpose of leveraging non-imaging information and fMRI \cite{parisot2018disease, parisot2017spectral, arya2020fusing}, have assumed the ability of GCN to be aware of inter-individual phenotypic differences and to regularize raw features based on the former. Even though the regularization phenomenon is proved, to some degree, as discussed above, there is no clear conclusion that the GCN really has learned the inter-individual non-imaging differences. To intuitively present the learned node embeddings, we have downsampled them onto the 2D plane with t-SNE proposed by \cite{van2008visualizing}. As shown in the left three plots of figure \ref{fig:2d}, even though graph pooling has detected some implicit inter-group heterogeneity, which is discussed in section \ref{sec:biomarker}, the features subsequently learned by MLP have not performed the relative feature distribution difference in respect of phenotypic information. The inevitable information loss during the feature dimensionality reduction may have caused this inconsistency, as the dimension of the basic features is up to 128. Still, the node embeddings learned by Graph Convolutional Networks have shown obvious clustering even in the 2D space. As exhibited in the right three diagrams of figure \ref{fig:2d}, the distance among subjects that are identical regarding a certain kind of phenotypic information is relatively close in the feature space compared to those who are not. This clustering is more evident when only considering the genders, which indicates that the learned edge weights of the population graph may depend more on it.

\section{Conclusion}
In this paper, we have proposed a framework to identify Autism Spectrum Disorders. First, we have proposed a novel self-attention downsampling method for fMRI. This end-to-end, unsupervised, and flexible graph pooling method has successfully considered brain functional connections and regional activities simultaneously, which can also be aware of individual differences in brain function. Besides, we have exploited Graph Convolutional Networks to incorporate imaging with phenotypic information and illustrated its efficiency in recalibrating the feature distribution. Our framework has achieved a mean accuracy of $87.62\%\pm0.06$ and a mean AUC of $0.92\pm0.01$ on ABIDE I dataset. The superior performance of our model indicates its ability to detect Autism Spectrum Disorders and contribute to early intervention.


\bibliographystyle{ieeetran} 
\bibliography{references} 

\end{document}